\documentclass[preprint,12pt]{elsarticle} 

\usepackage{lineno}
\modulolinenumbers[5]
\usepackage{color}
\usepackage[usenames,dvipsnames]{xcolor}
\usepackage{fullpage}
\usepackage{amsmath}
\usepackage{amssymb}
\usepackage{mathrsfs}
\usepackage{newfloat}
\usepackage{graphicx}
\usepackage{subfig}
\usepackage{multirow}
\usepackage{stmaryrd}
\usepackage{caption}
\usepackage{tikz}
\usepackage{multicol}
\usetikzlibrary{arrows,positioning,calc}
\usepackage{pstool}
\usepackage{textcomp}
\usepackage[hidelinks]{hyperref}
\hypersetup{colorlinks,bookmarksopen,linkcolor=blue,citecolor=blue,urlcolor=blue}

\definecolor{myblue}{RGB}{0,115,189}
\definecolor{mygreen}{rgb}{0.19,0.61,0.21}

\newcommand{\bs}[1]{{\boldsymbol{#1}}}
\newcommand{\tr}[1]{\mathrm{tr\,{#1}}}

\DeclareFloatingEnvironment{algorithm}
\graphicspath{{figs/}}

\journal{Extreme Mech. Lett.}

\biboptions{authoryear}

\begin{document}

\begin{frontmatter}

\title{Extended Micromorphic Computational Homogenization for Mechanical Metamaterials Exhibiting Multiple Geometric Pattern Transformations\tnoteref{mytitlenote}}
\tnotetext[mytitlenote]{The post-print version of this article is published in \emph{Extreme Mech. Lett.}, \href{https://doi.org/10.1016/j.eml.2020.100708}{10.1016/j.eml.2020.100708}}

\author[TUe]{O.~Roko\v{s}\corref{correspondingauthor}} 
\ead{O.Rokos@tue.nl}

\author[TUe]{M.M.~Ameen}
\ead{M.Ameen@tue.nl}

\author[TUe]{R.H.J.~Peerlings}
\ead{R.H.J.Peerlings@tue.nl}

\author[TUe]{M.G.D.~Geers}
\ead{M.G.D.Geers@tue.nl}

\address[TUe]{Mechanics of Materials, Department of Mechanical Engineering, Eindhoven University of Technology, P.O.~Box~513, 5600~MB~Eindhoven, The~Netherlands}
\cortext[correspondingauthor]{Corresponding author.}

\begin{abstract}
	
Honeycomb-like microstructures have been shown to exhibit local elastic buckling under compression, with three possible geometric buckling modes, or pattern transformations. The individual pattern transformations, and consequently also spatially distributed patterns, can be induced by controlling the applied compression along two orthogonal directions. Exploitation of this property holds great potential in, e.g., soft robotics applications. For fast and optimal design, efficient numerical tools are required, capable of bridging the gap between the microstructural and engineering scale, while capturing all relevant pattern transformations. A micromorphic homogenization framework for materials exhibiting multiple pattern transformations is therefore presented in this paper, which extends the micromorphic scheme of~\citeauthor{Rokos2019}, \emph{J. Mech. Phys. Solids}~{\bf 123}, 119--137 (2019), for elastomeric metamaterials exhibiting only a single pattern transformation. The methodology is based on a suitable kinematic ansatz consisting of a smooth part, a set of spatially correlated fluctuating fields, and a remaining, spatially uncorrelated microfluctuation field. Whereas the latter field is neglected or condensed out at the level of each macroscopic material point, the magnitudes of the spatially correlated fluctuating fields emerge at the macroscale as micromorphic fields. We develop the balance equations which these micromorphic fields must satisfy as well as a computational homogenization approach to compute the generalized stresses featuring in these equations. To demonstrate the potential of the methodology, loading cases resulting in mixed modes in both space and time are studied and compared against full-scale numerical simulations. It is shown that the proposed framework is capable of capturing the relevant phenomena, although the inherent multiplicity of solutions entails sensitivity to the initial guess.

\end{abstract}

\begin{keyword}
Computational homogenization \sep pattern transformation \sep micromorphic continuum \sep non-linear homogenization \sep mechanical metamaterials \sep elastomeric honeycombs
\end{keyword}

\end{frontmatter}


%
%
\section{Introduction}
\label{introduction}
\begin{figure}
	\centering
	\subfloat[square pattern]{\includegraphics[height=4cm]{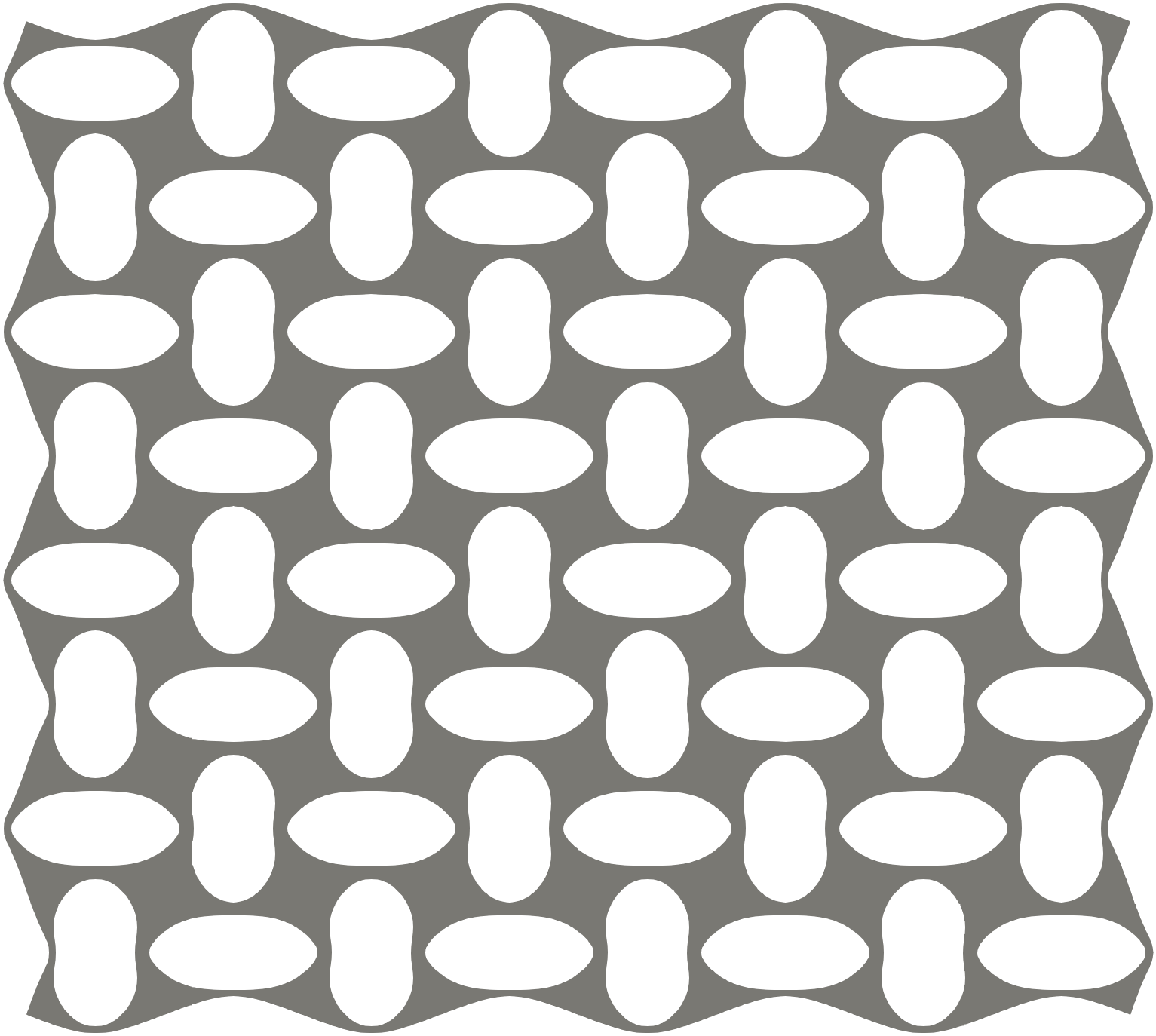}\label{Figure.Pattern_1a}}
	\hspace{0.25em}
	\subfloat[hexagonal pattern~I]{\includegraphics[height=4cm]{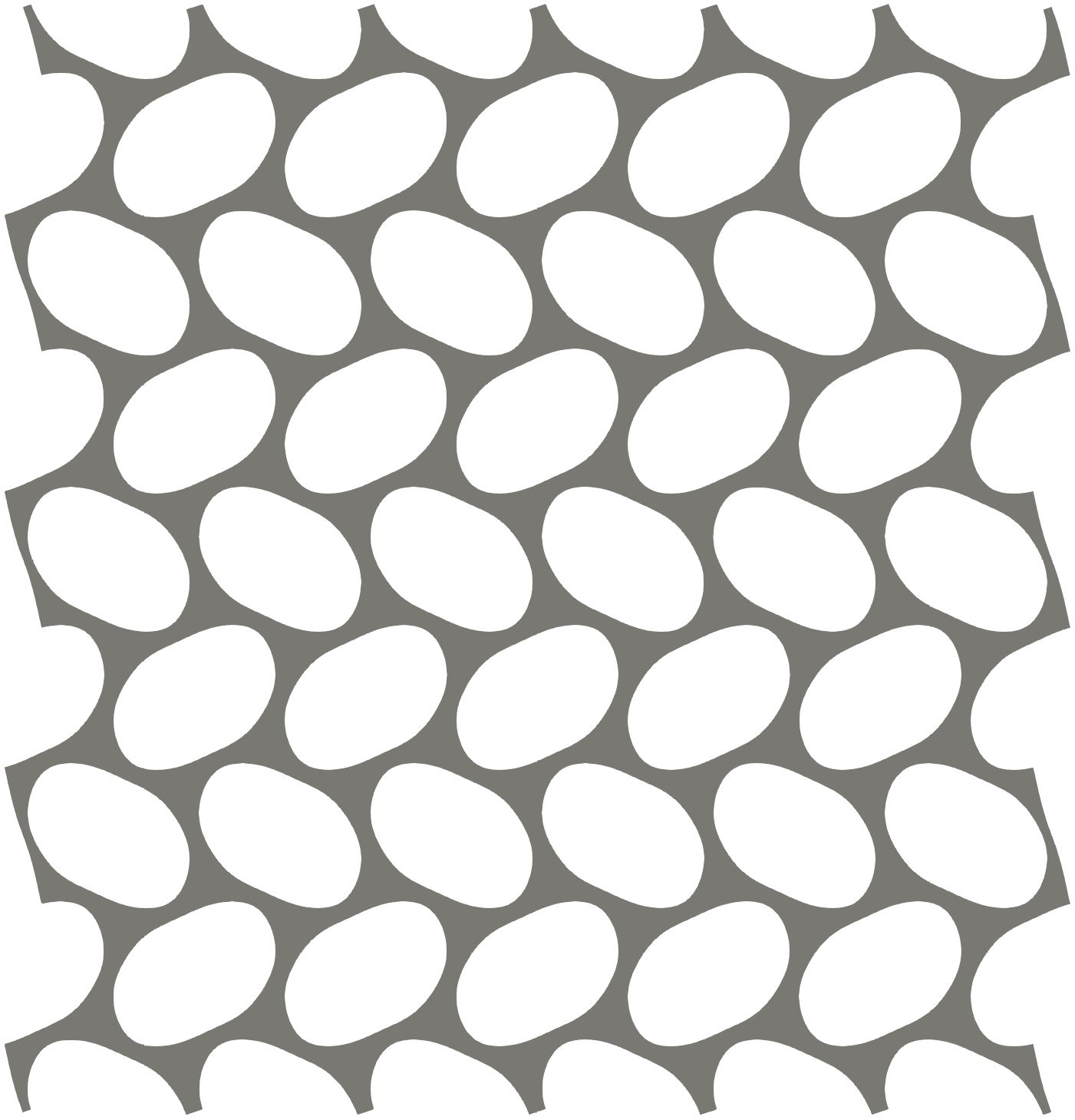}\label{Figure.Pattern_1b}}
	\hspace{0.25em}
	\subfloat[hexagonal pattern~II]{\includegraphics[height=4cm]{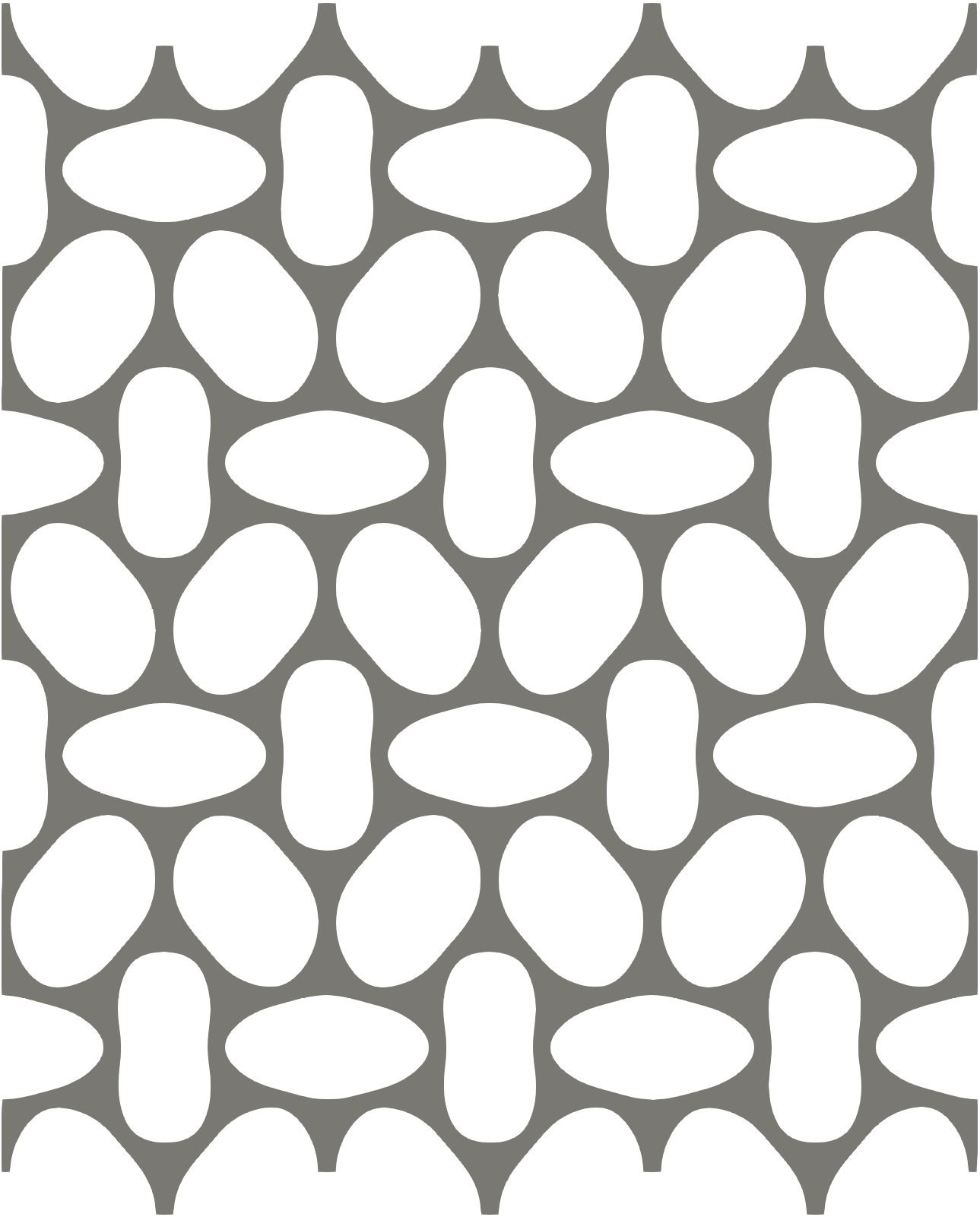}\label{Figure.Pattern_1c}}
	\hspace{0.25em}
	\subfloat[hexagonal pattern~III]{\includegraphics[height=4cm]{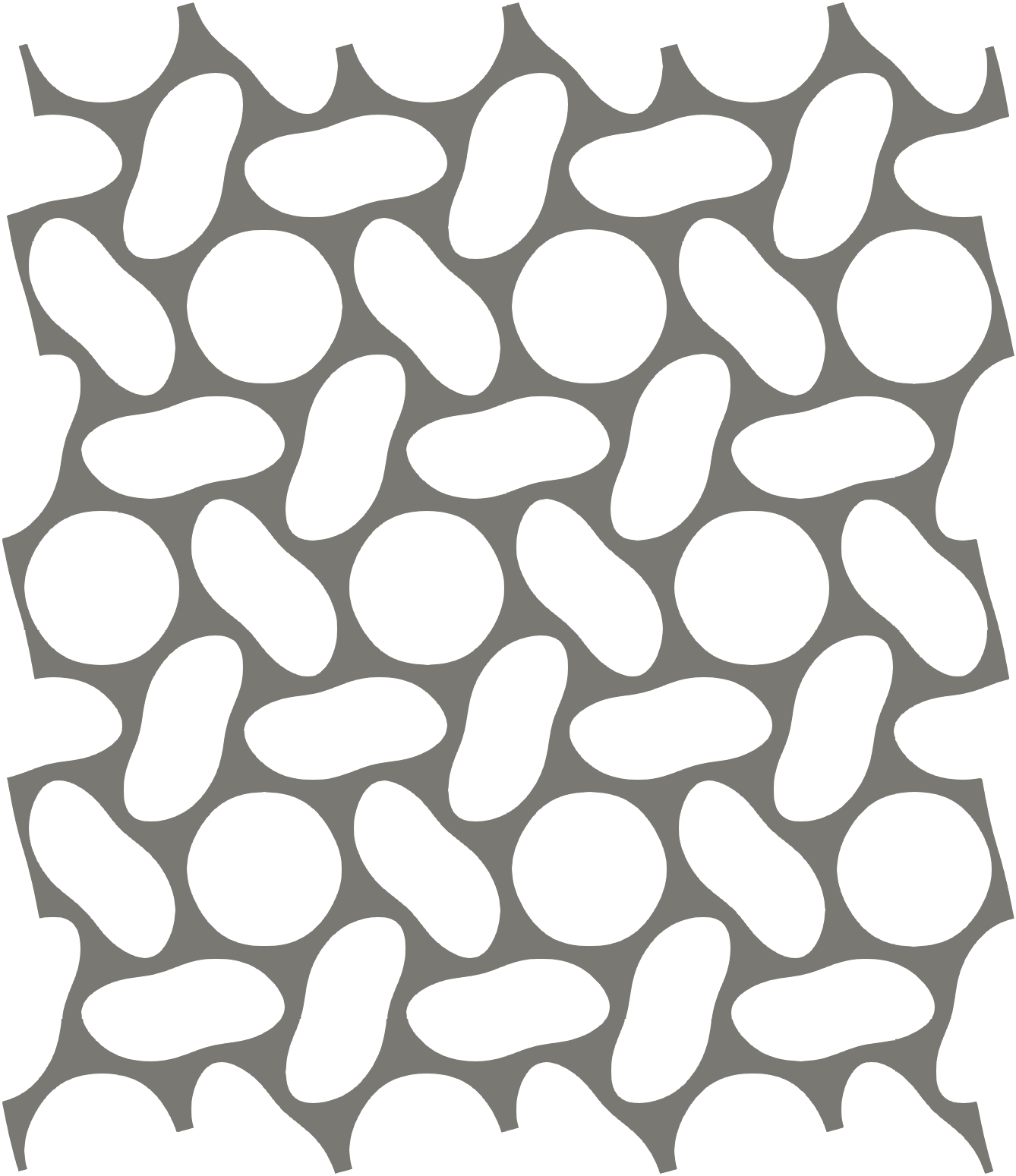}\label{Figure.Pattern_1d}}
	\caption{Typical geometric pattern transformations exhibited by voided elastomeric materials. (a)~Single pattern transformation for a square packing of holes (uniaxial or biaxial compression). Multiple pattern transformations exhibited by microstructures with hexagonal packing of holes: (b)~\emph{pattern~I} or \textit{shear pattern} (uniaxial compression), (c)~\emph{pattern~II} or \textit{butterfly-like pattern} (biaxial compression with a higher compression along the horizontal axis), and~(d) \emph{pattern~III} or \textit{flower-like pattern} (equi-biaxial compression).}
	\label{introduction:fig1}
\end{figure}

Elastomeric materials with specific microstructures, typically consisting of periodically arranged holes, have been shown to reveal geometric pattern transformations under compression, see Fig.~\ref{Figure.Pattern_1a}, as well as e.g.~\cite{Mullin2007a}, \cite{Bertoldi:2010a}, or~\cite{florijn2014}. The pattern transformation is a result of ordered local buckling of the elastomeric cell walls, resulting in a distinct post-bifurcation behaviour, which characterizes such materials as mechanical metamaterials. The post-bifurcation compliance has been shown to make these materials especially suitable for soft robotics applications~\citep[see e.g.][]{Whitesides:2018}.

Of particular interest in this contribution are elastomeric materials exhibiting multiple pattern transformations, depending on the loading conditions. Honeycomb microstructures, for instance, show three distinct patterns under biaxial compression, as depicted in Figs.~\ref{Figure.Pattern_1b}--\ref{Figure.Pattern_1d}, where a periodic cell with~$ 6 \times 8 $ holes is shown. Such patterns have been observed both experimentally and numerically~\citep[cf.][]{Gibson1999,Gibson1989,Papka1999a,Papka1999b,Guo1999,Chung1999, Ohno2002,Karagiozova2008,Haghpanah2014,Shan2014a,Combescure2016}. In particular, \cite{Ohno2002} studied symmetric bifurcations in honeycomb microstructures and showed that it is the multiplicity of the bifurcation (double and triple) that leads to complex patterned modes. This implies the existence of multiple independent buckling modes, any linear combination of which can result in an admissible pattern in an infinite medium. \cite{Okumura2002} furthermore studied the post-bifurcation behaviour of these microstructures under equi-biaxial loading conditions, to understand which of the possible microscopic buckling mode combinations are preferential based on the associated elastically stored energies.

\begin{figure}
	\centering
	\subfloat[square pattern]{\includegraphics[height=4cm]{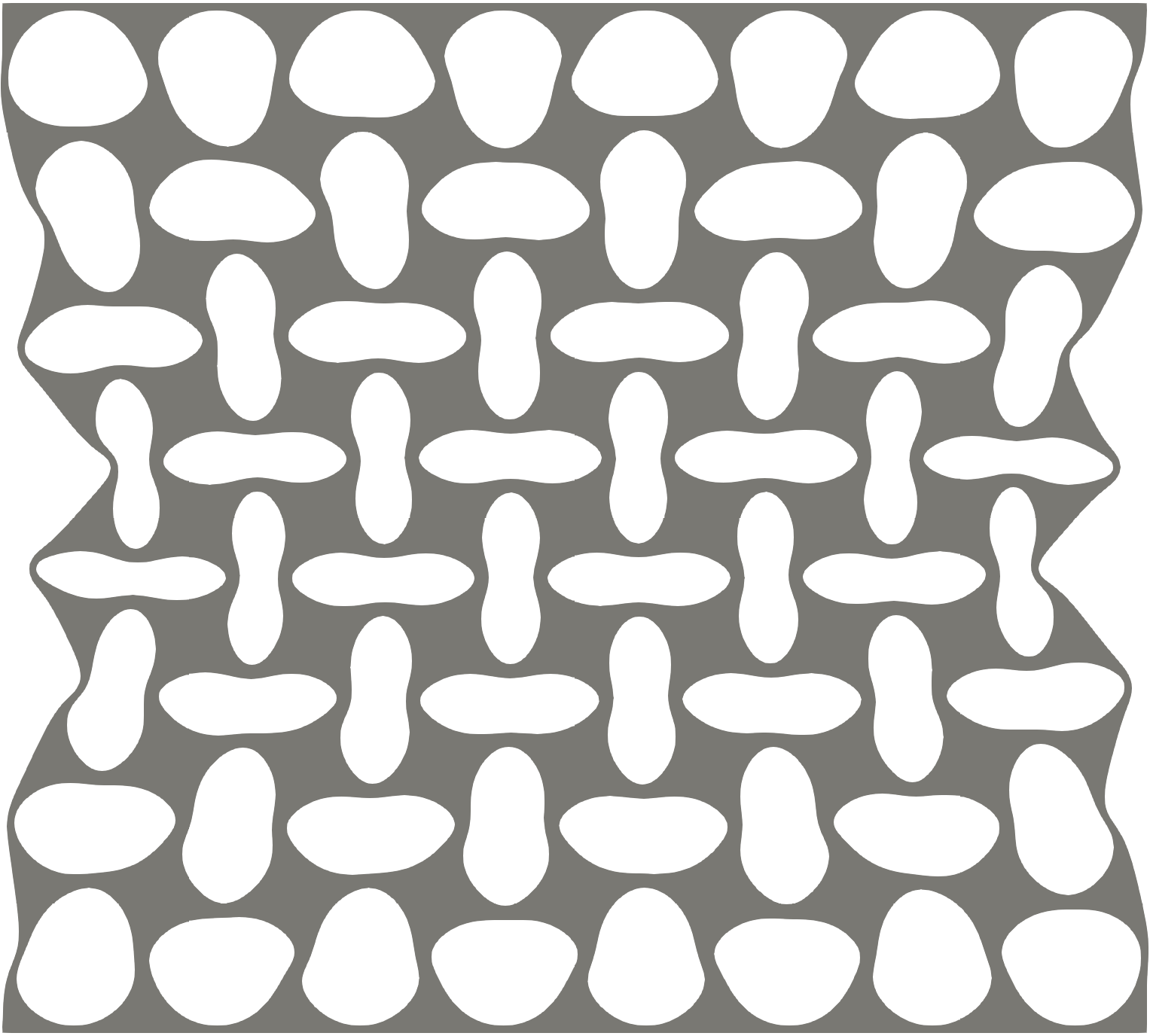}\label{Figure.BL_1a}}\hspace{0.25em}
	\subfloat[hexagonal pattern~I]{\includegraphics[height=4cm]{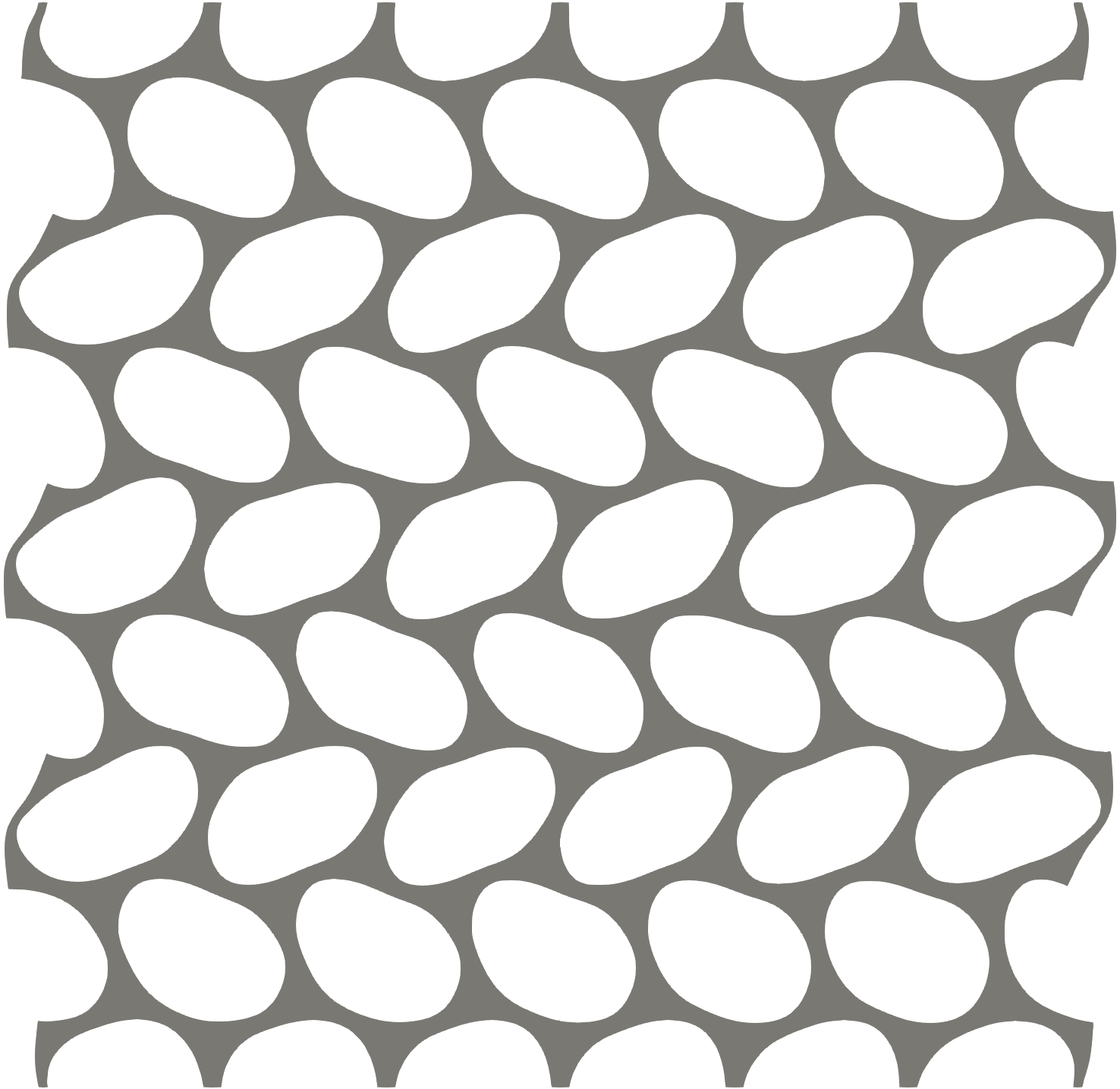}\label{Figure.BL_1b}} \hspace{0.25em}
	\subfloat[hexagonal pattern~II]{\includegraphics[height=4cm]{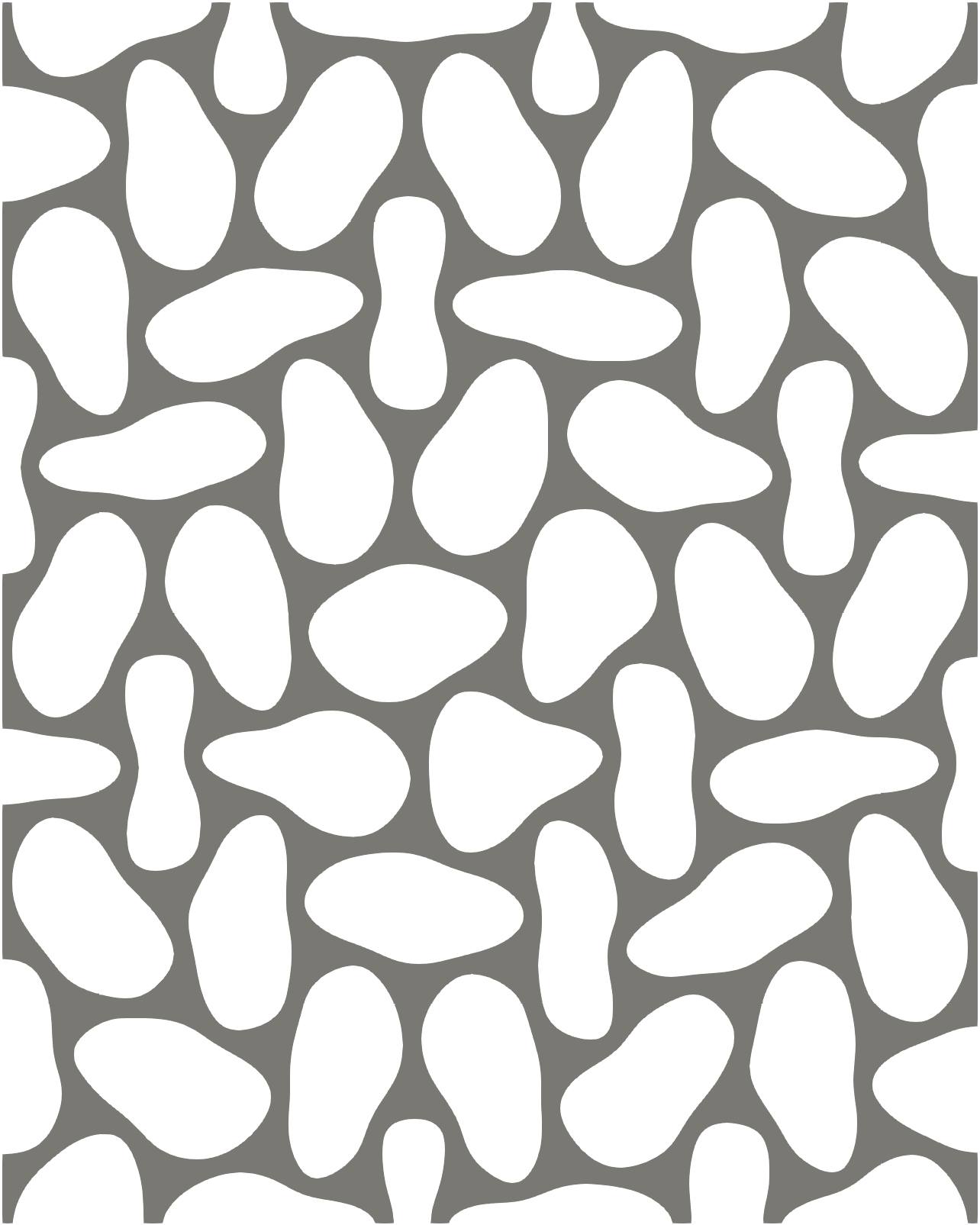}\label{Figure.BL_1c}}\hspace{0.25em}
	\subfloat[spatial mixing of patterns]{\includegraphics[height=4cm]{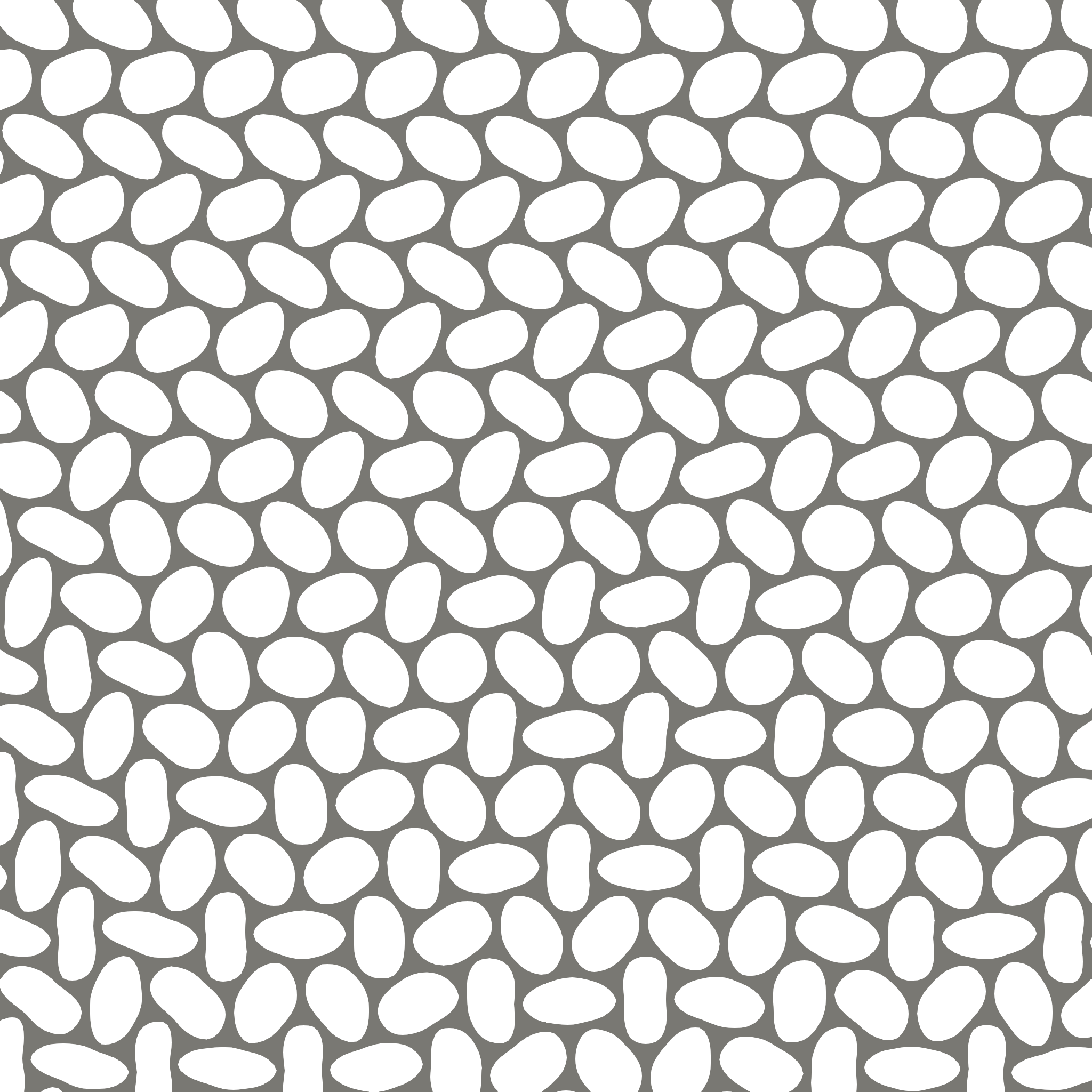}\label{Figure.BL_1d}}	
	\caption{Boundary layers exhibited by finite specimens for various patterns. (a)~A square specimen with an~$ 8 \times 8 $ square grid of holes subjected to uniaxial vertical compression of~$10\%$ with clamped top and bottom edges, whereas the lateral edges are unconstrained. (b)~Uniaxial $5\%$ vertical compression of hexagonally stacked microstructure with~$6$ and~$8$ holes along the horizontal and vertical directions, and with clamped top and bottom edges, and unconstrained vertical edges. (c)~Biaxial compression of hexagonally stacked microstructure subjected to~$2.5\%$ vertical and~$5\%$ horizontal strain. (d)~A case of spatial mixing of patterns, showing all the three patterns~I, II, and III, under complex loading and boundary conditions.}
	\label{BL:fig6}
\end{figure}

For finite specimens, boundary layers are formed where kinematic constraints exerted by essential boundary conditions restrict the pattern transformations. As shown by~\cite{Ameen:MM}, the relative thickness of the boundary layers with respect to the overall size of the specimen introduces a size effect, which may significantly influence the overall behaviour of the specimen. This effect is strong especially for small ratios of the typical macroscopic size~$L$ (e.g.~specimen size) and the size of typical microstructural features~$\ell$ (e.g.~hole spacing), which is generally called the scale ratio. Figs.~\ref{Figure.BL_1a}--\ref{Figure.BL_1c} show several examples of typical boundary layers formed in the vicinity of kinematically restricted boundaries. They comprise approximately~$3$ layers of holes close to the top and bottom edges in the case of square-packed microstructure (Fig.~\ref{Figure.BL_1a}), whereas in the case of a hexagonally-packed microstructure the observed thicknesses are smaller (Figs.~\ref{Figure.BL_1b} and~\ref{Figure.BL_1c}). As a consequence of the external loading, individual patterns may in addition alter and switch in time, or mix in space, as shown in Fig.~\ref{Figure.BL_1d}. Such a behaviour clearly poses a challenging problem from a homogenization perspective.

In the limit of large scale separation, conventional computational homogenization methods provide accurate predictions even for very complex microstructural behaviour \cite[cf.~e.g.][]{Kouznetsova:2001,Miehe:2002}. Yet, they fail in the regime of low scale separation, where individual cells are strongly kinematically coupled. To capture such mutual interactions and the ensuing non-locality, several higher-order homogenization schemes that incorporate long-range fluctuation fields have been proposed in the literature~\cite[e.g.][]{Kouznetsova:2004}. Such frameworks typically result in a higher-order macroscopic continuum. Another class consists of the generalized micromorphic theories, resulting in an extended continuum at the macroscale, such as micromorphic homogenization schemes proposed by~\cite{Hutter:2017} or~\cite{Biswas:2017}. A micromorphic computational homogenization scheme specifically targeting mechanical metamaterials exhibiting a single pattern transformation (see Fig.~\ref{Figure.Pattern_1a}) has been recently proposed in~\citep{Rokos2019}. A unique decomposition of the underlying kinematic field is adopted to incorporate the spatially correlated microfluctuating component in addition to the mean displacement field and the uncorrelated local microfluctuation field. This long-range correlated field directly relates to the microstructural morphology, and accounts for the kinematical interaction between individual microstructural periodic cells. It has been shown that such a single-pattern computational scheme is able to capture the ensemble average of the full-scale results and the size effects exhibited by them, see~\cite{Rokos2019}.

It is the objective of this paper to formulate a homogenization framework that captures multiple pattern transformations, including their spatial and temporal mixing as well as associated boundary layers. To this end, the micromorphic homogenization scheme of~\cite{Rokos2019} is extended to the case of multiple pattern transformations. This is not a trivial extension given that multiplicity of bifurcation modes can occur, and that individual modes may mutually interact and evolve. Although the presented framework is fully general, we will use honeycomb-like microstructures with a hexagonal stacking of holes to guide our developments and to illustrate potential of the method. To this end, the fundamental patterns of a hexagonal microstructure are first identified and computed~\cite[based on][]{Ohno2002}, which are later used to define the long-range spatially correlated fields. The amplitudes of these fields are controlled by scalar and slowly varying micromorphic fields, which quantify the presence of the patterns as a consequence of external loading. By using variational principles, the corresponding macroscopic governing equations are derived. The locally uncorrelated microfluctuation field is localized using a periodicity assumption and is condensed out at the level of each macroscopic Gauss integration point. Because, as demonstrated in our earlier work, the micromorphic framework was specifically designed to capture size effects resulting from finite specimens with small scale separation, the examples presented here focus on the novel aspect of this paper, i.e.~on the ability to deal with multiple patterns and transitions between them---in time and space.

The remainder of this paper is organized as follows. Section~\ref{problem} introduces the considered microstructure, corresponding periodic cell, and individual geometric patterns and modes. Section~\ref{homogenization} extends the micromorphic homogenization framework to the case of multiple pattern transformations and provides the resulting macroscopic governing equations. Section~\ref{results} then compares the homogenization results against full-scale Direct Numerical Simulation~(DNS) on two examples of temporal switching and spatial mixing of modes. The paper closes with a summary and conclusions in Section~\ref{summary}.

The following notation conventions are used throughout this manuscript
\begin{multicols}{2}
\begin{itemize}[\textbf{-}]
\itemsep0em 
\item scalars~$ a $,
\item vectors~$ \vec{a} $,
\item position vector~$ \vec{X} = X_1\vec{e}_1 + X_2\vec{e}_2 $,
\item second-order tensors~$ \bs{A} $,
\item matrices~$ \bs{\mathsf{A}} $ and column matrices~$ \underline{a} $,
\item $ \vec{a} \cdot \vec{b} = a_i b_j $,
\item $ \bs{A} \cdot \vec{b} =  A_{ij} b_j\vec{e}_i $,
\item $\bs{A}\cdot\bs{B} = A_{ik}B_{kj}\vec{e}_i\vec{e}_j $,
\item $\bs{A}:\bs{B} = A_{ij}B_{ji}$,
\item transpose~$ \bs{A}^\mathsf{T}$, $ A_{ij}^\mathsf{T} = A_{ji} $,
\item gradient~$ \displaystyle \vec{\nabla} \vec{a}(\vec{X}) = \frac{\partial a_j(\vec{X})}{\partial X_i} \vec{e}_i \vec{e}_j $,
\item divergence~$ \displaystyle \vec{\nabla} \cdot \vec{a}(\vec{X}) = \frac{\partial a_i(\vec{X})}{\partial X_i} $,
\item derivatives of scalar functions with respect to second-order tensors \\
$\displaystyle \delta\Psi(\bs{F};\delta\bs{F}) = \left.\frac{\mathrm{d}}{\mathrm{d}h}\Psi(\bs{F}+h\delta\bs{F})\right|_{h=0} = \frac{\partial\Psi(\bs{F})}{\partial\bs{F}}:\delta\bs{F} $,
\end{itemize}
\end{multicols}
\noindent
where~$\vec{e}_i$, $i = 1, 2,$ are basis vectors of a chosen coordinate frame, and where Einstein's summation convention is assumed on repeated indices~$i$, $j$, or~$k$.
%
%
\section{Problem Description}
\label{problem}
%
%
\subsection{Material Properties, Microstructural Geometry, and Periodic Cell}
\label{micro_prop}
The constitutive behaviour of the elastomeric base material is assumed to be hyperelastic, described by the following strain energy density
\begin{equation}
\psi\left(\bs{F}\right) = a_1 (I_1 - 3) + a_2 (I_1 - 3)^2 - 2 a_1 \log{J} + \frac{1}{2} K (J-1)^2,
\label{problem:eq1}  
\end{equation}
where~$ \bs{F} = \bs{I} + ( \vec{\nabla} \vec{u} )^\mathsf{T} $ is the deformation gradient tensor, $\vec{\nabla}\vec{u}$ is the gradient of the displacement field with respect to the reference configuration, $\bs{I}$ is the second-order identity tensor, $ J $ is the determinant of~$ \bs{F} $, and~$I_1 = \tr{\bs{C}}$ is the first invariant of the right Cauchy--Green deformation tensor~$ \bs{C} = \bs{F}^\mathsf{T} \cdot \bs{F} $. The adopted material parameters are given by~$ a_1 = 0.55~$~MPa, $ a_2 = 0.3$~MPa, and~$K = 55$~MPa~\citep[cf.][]{Bertoldi2008d}.

\begin{figure}
	\centering
	\subfloat[periodic cell~$Q$]{\def\svgwidth{3.75cm}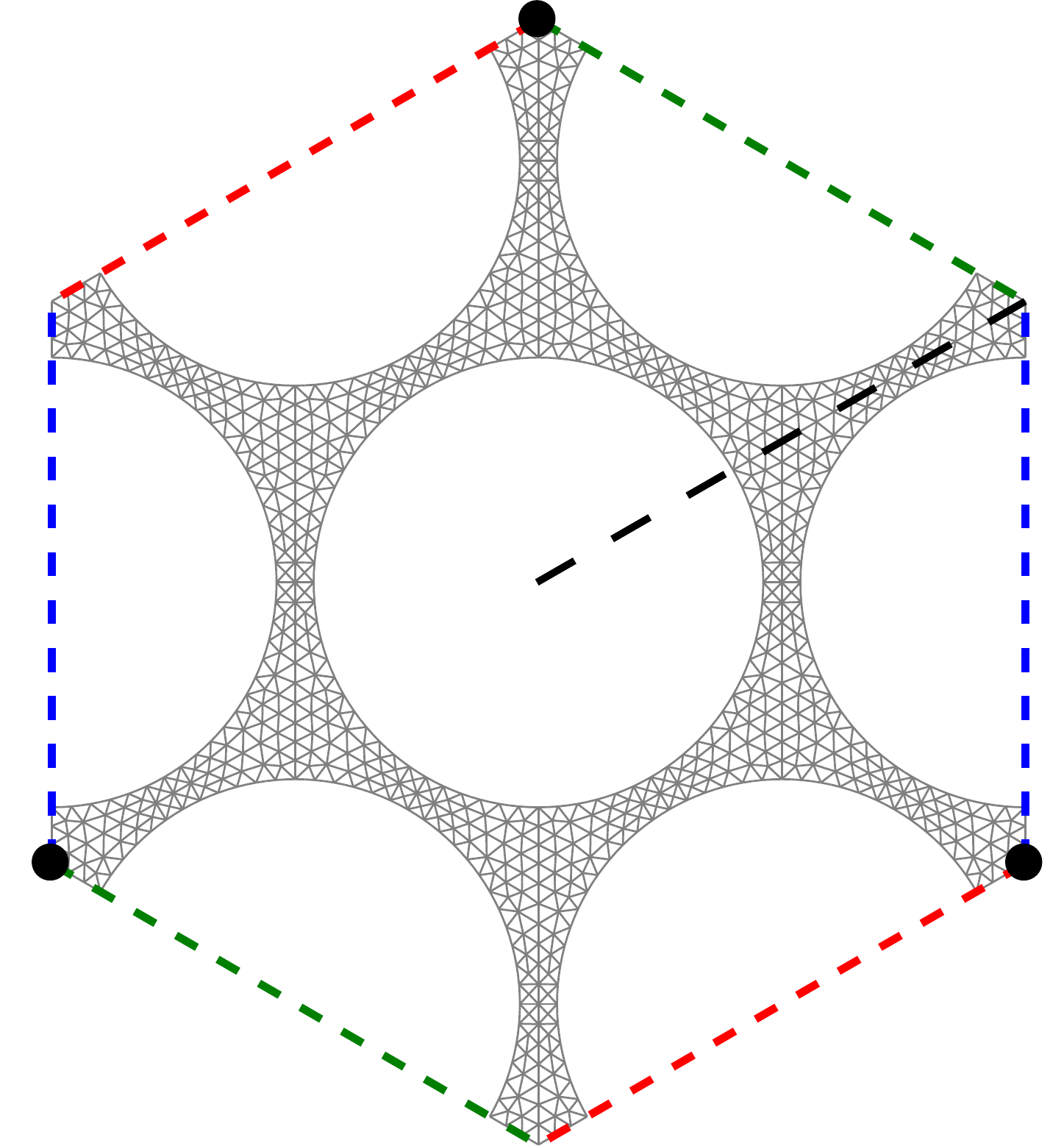\label{problem:fig2a}}\hspace{0.25em}
	\subfloat[pattern~I]{\includegraphics[height=4.0cm]{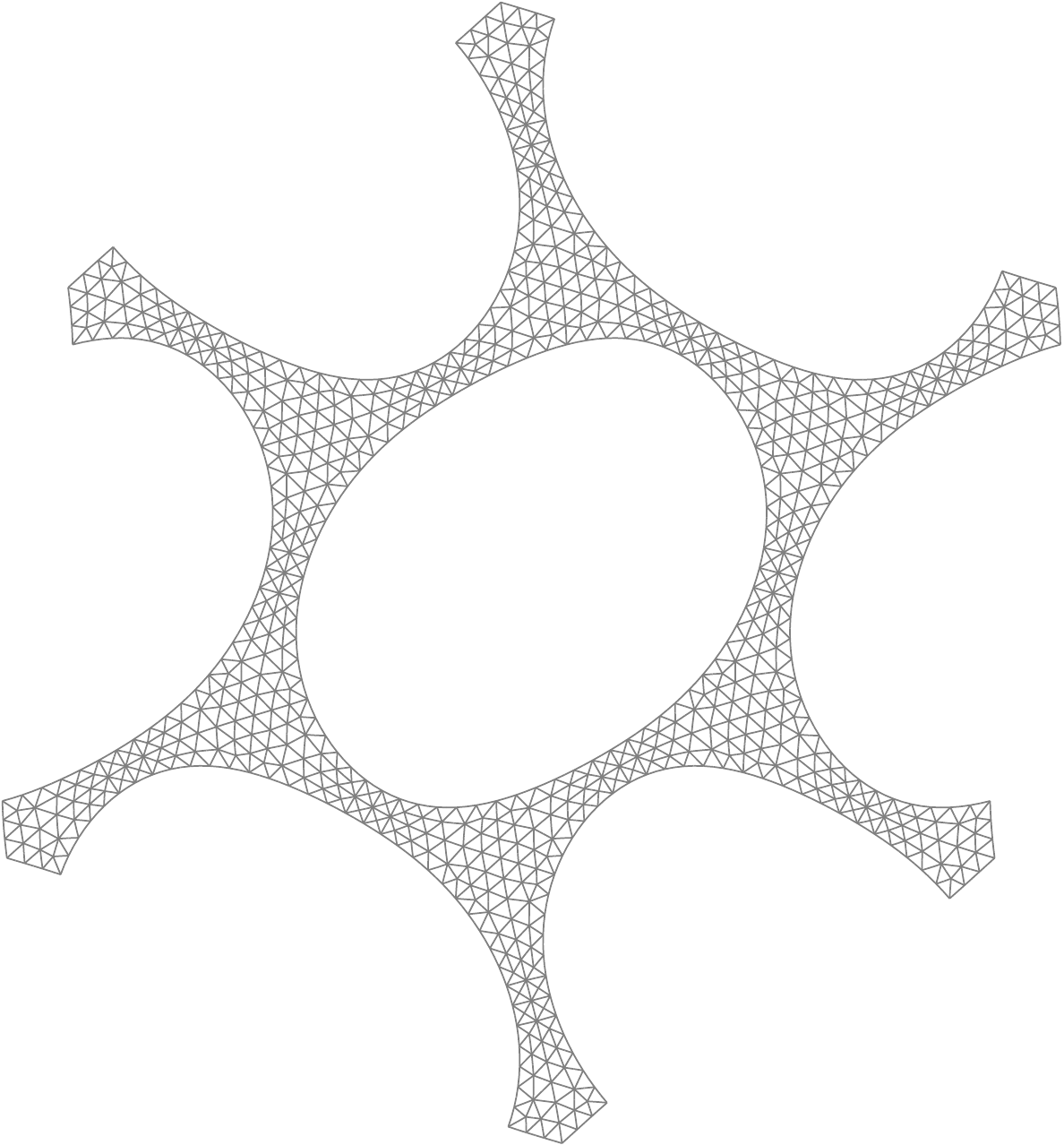}\label{problem:fig2b}}\hspace{0.25em}
	\subfloat[pattern~II]{\includegraphics[height=4.0cm]{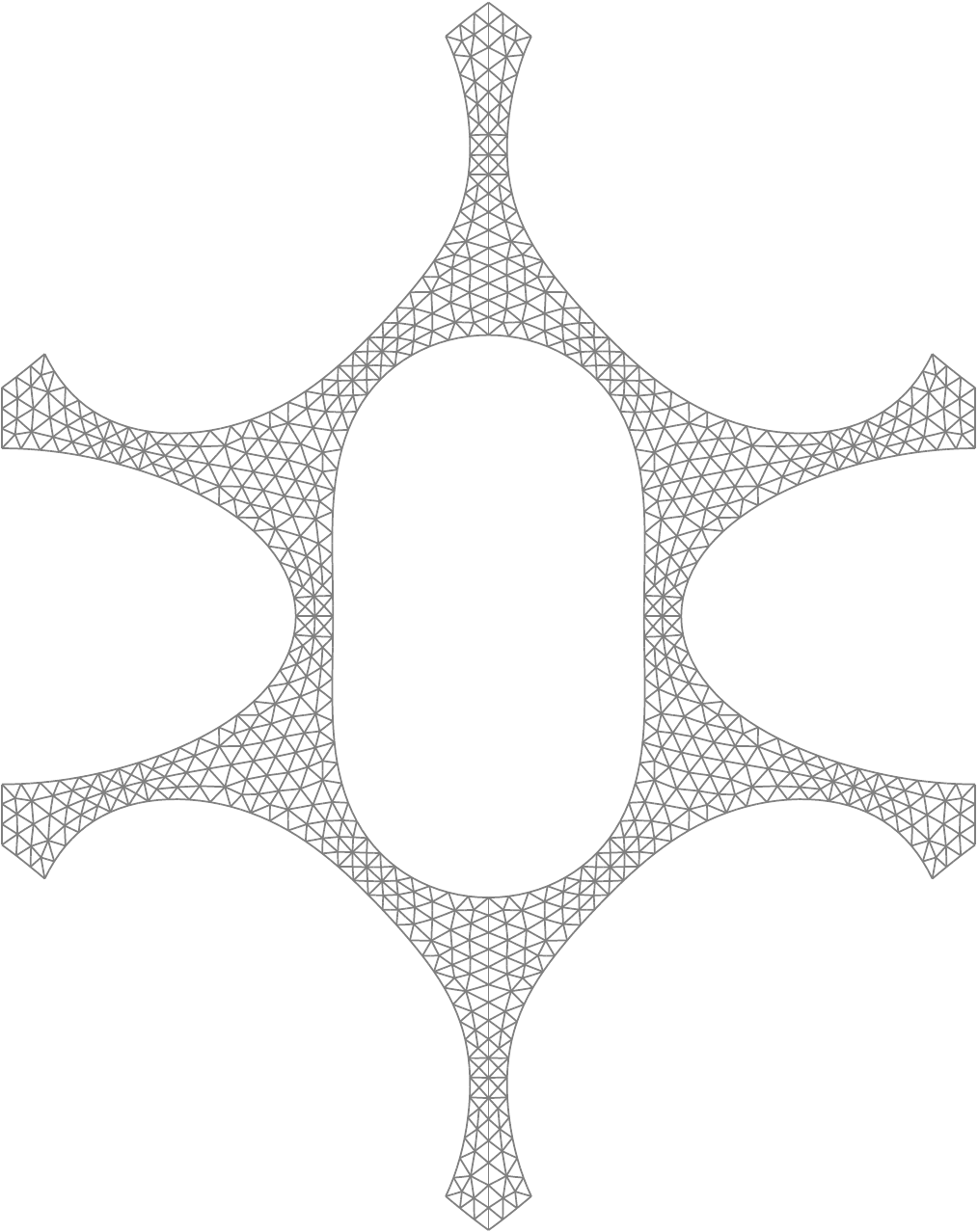}\label{problem:fig2c}}\hspace{0.25em}
	\subfloat[pattern~III]{\includegraphics[height=4.0cm]{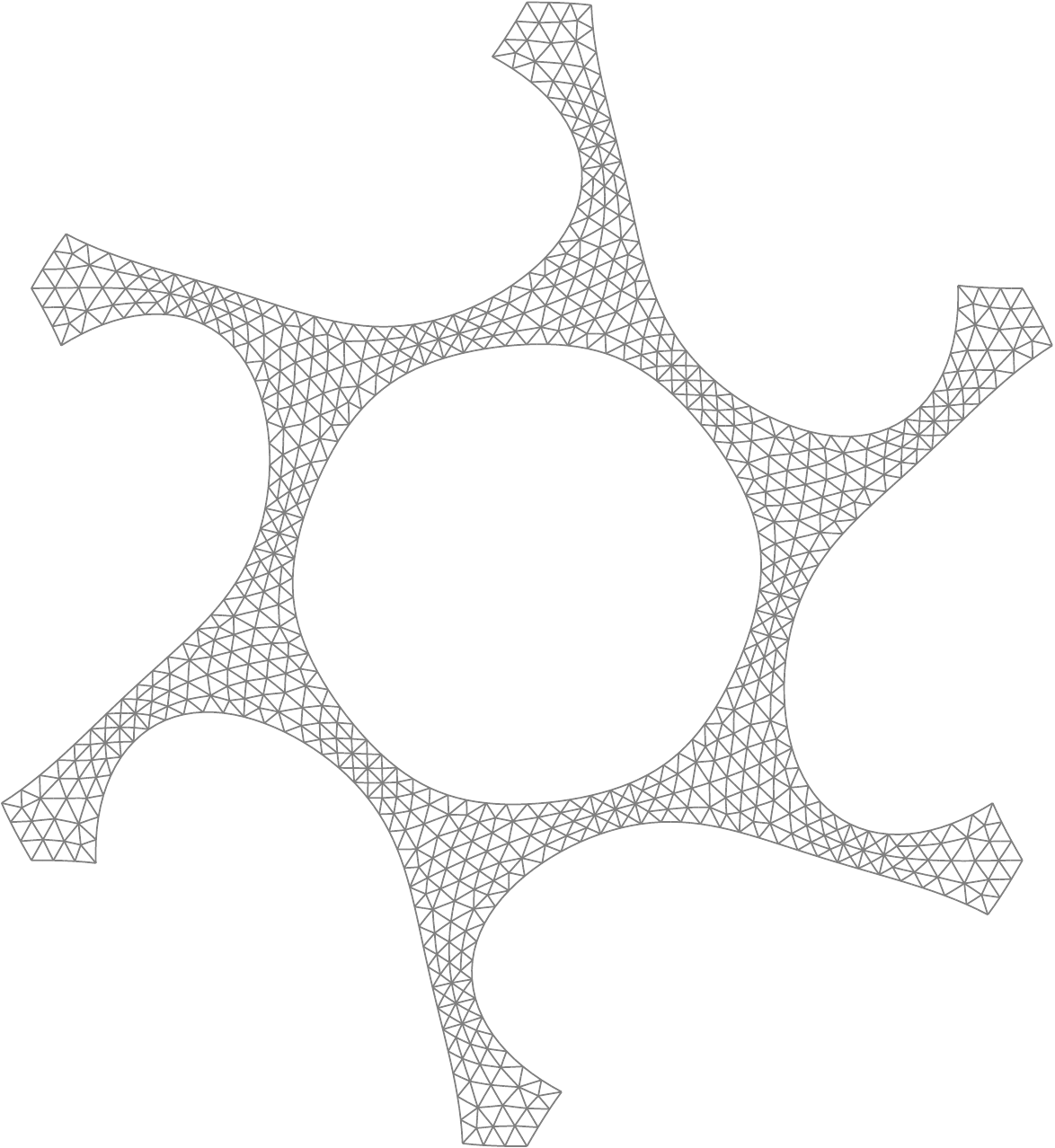}\label{problem:fig2d}}
	\caption{(a)~Adopted periodic cell~$Q$ for the elastomeric microstructure with a hexagonal stacking of holes. Three sets of edges are periodically tied according to their colour coding (green, blue, and red). The overall strain is imposed through a set of three control points, namely~$P_1$, $P_2$, and $P_3$ (black dots). Pattern transformations corresponding to (b)~pattern~I, uniaxial pattern, or shear pattern, $\vec{\pi}_1$ (computed for~$\gamma = \infty$), (b)~pattern~II, biaxial pattern, or butterfly-like pattern, $\vec{\pi}_2$ (computed for~$\gamma = \frac{1}{2}$), and~(c) pattern~III, equi-biaxial pattern, or butterfly-like pattern, $\vec{\pi}_3$ (computed for~$\gamma = 1$).}
	\label{problem:fig2}
\end{figure}

A hexagonal stacking of circular holes with diameter~$d=1.28$~mm and a centre-to-centre spacing of $\ell=1.386$~mm is adopted, representing the microstructural morphology. The periodic cell~$Q$ considered is depicted in Fig.~\ref{problem:fig2a}. It comprises~$ 2 $ holes in each direction of the cell walls and is consistent with the periodicity of the observed patterns~\citep[see e.g.][]{Bertoldi2008d,Ameen:MM,Papka1999a,Papka1999b,Ohno2002}. The periodic cell is fully symmetric with respect to the adopted Cartesian coordinate frame. Periodicity is enforced by tying conditions highlighted by colour coding, and the cell's macroscopic deformation is controlled through a set of three points, as shown in Fig.~\ref{problem:fig2a}.
%
%
\subsection{Pattern Transformations and Modes}
\label{RVE_modes}
Various overall biaxial compressive strain paths are applied. They are characterized by a biaxiality ratio~$\gamma$, defined as
\begin{equation}
\gamma = \frac{|\overline{F}_{22}-1|}{|\overline{F}_{11}-1|} = \frac{|\varepsilon_{22}|}{|\varepsilon_{11}|} \in [ 0, \infty ],
\label{biaxiality}
\end{equation}
where~$\varepsilon_{ii} = \overline{F}_{ii}-1$ are the normal compressive strains along the~$\vec{e}_1$ and $\vec{e}_2$ directions, $\overline{F}_{12} = \overline{F}_{21} = 0$, and $\overline{F}_{ij}$ denotes the components of the overall deformation gradient tensor~$\overline{\bs{F}}$. Depending on the value of~$\gamma$, various patterns can be triggered in the adopted elastomeric microstructure. Using the classification by~\cite{Ohno2002}, the three directions of the cell walls (denoted as~$\theta = 90^\circ$ and~$ \pm 30^\circ$, where~$\theta$ is defined with respect to the horizontal axis at the intersection of the cell walls, cf. Fig.~\ref{problem:fig2a}) are used to distinguish individual patterns as follows:
\begin{enumerate}[(i)]\setlength{\itemsep}{0pt}\setlength{\parskip}{0pt}\setlength{\parsep}{0pt}

	\item \emph{Pattern~I}, also called \emph{uniaxial pattern}, or \textit{shear pattern}, is observed for the loading cases in which the cell walls along~$\theta = 90^\circ$ take a higher compressive load compared to the remaining cell walls (i.e.~$\gamma > 1$), and a single bifurcation point occurs. The displacement field associated with this pattern is denoted~$\vec{\pi}_1(\vec{X})$, and corresponds to alternating horizontal layers of holes buckled towards the right and left with no auxetic effect, cf. Figs.~\ref{Figure.Pattern_1b} and~\ref{problem:fig2b}.
	
	\item \emph{Pattern~II}, a \emph{biaxial pattern}, or \textit{butterfly-like pattern}, is observed when the cell walls along~$\theta = \pm 30^\circ$ take an equal compressive load which is higher than that of the cell walls along~$\theta = 90^\circ$ (i.e.~$\gamma < 1$), and a double bifurcation point occurs. This pattern is denoted~$\vec{\pi}_2(\vec{X})$, and corresponds to one layer of holes flattened alternately along the horizontal and vertical directions, whereas the adjacent layers of holes buckle alternately towards the left and right, forming a butterfly-like pattern, cf. Figs.~\ref{Figure.Pattern_1c} and~\ref{problem:fig2c}.
		
	\item \emph{Pattern~III}, an \emph{equi-biaxial pattern}, or \textit{flower-like pattern}, emerges when all three cell walls are subjected to an equal compressive load, occurring only for the case of equi-biaxial compression (i.e.~$ \gamma = 1 $), in which a triple bifurcation point occurs. This pattern is denoted~$\vec{\pi}_3(\vec{X})$, and corresponds to a virtually undeformed hole surrounded by ellipses with their major axes positioned tangentially to that hole, cf. Figs.~\ref{Figure.Pattern_1d} and~\ref{problem:fig2d}.

\end{enumerate}

\begin{figure}
      \centering
      \subfloat[periodic cell~$Q$]{\includegraphics[height=4.0cm]{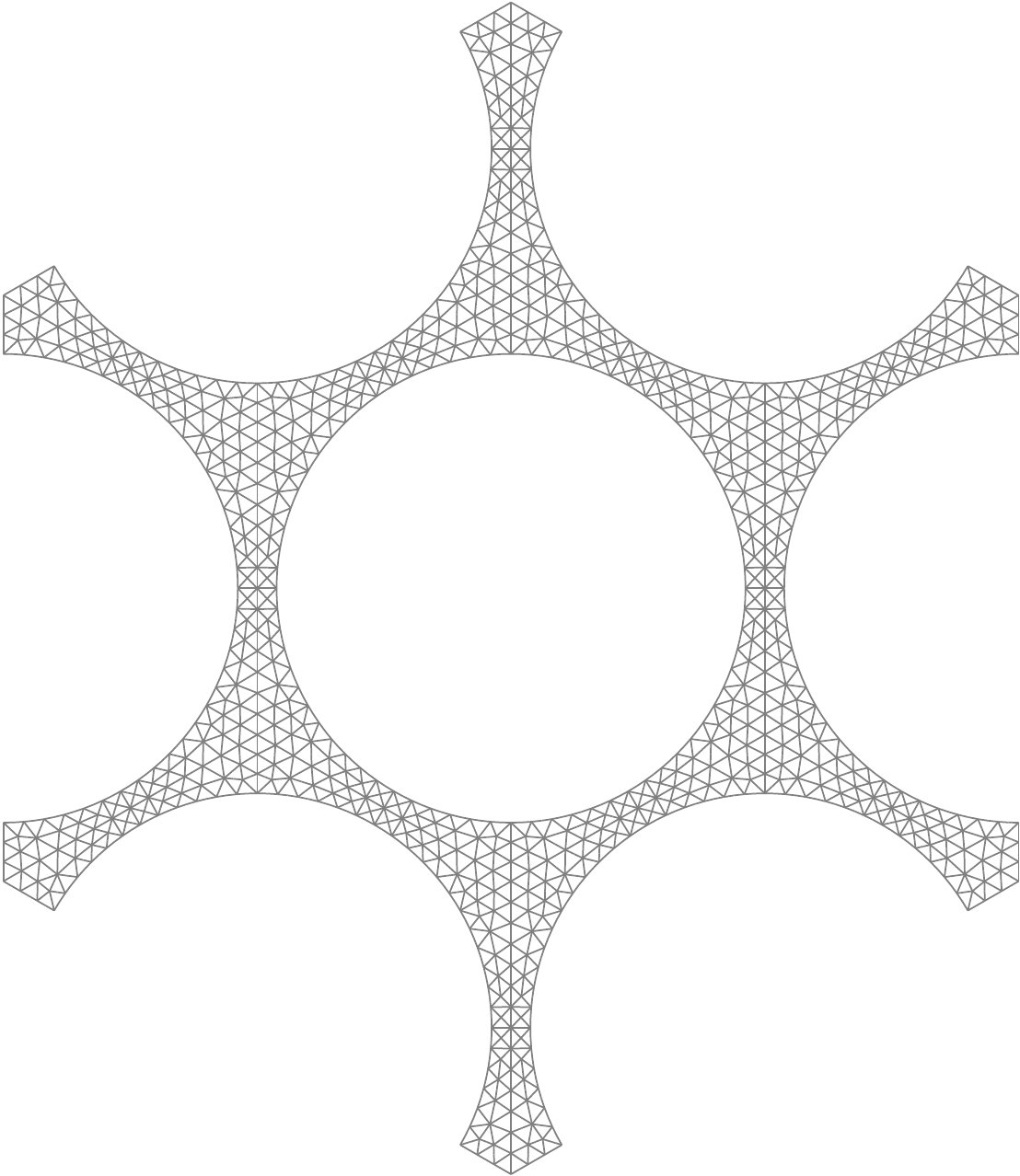}\label{Figure.sketch_3a}}\hspace{0.25em}
      \subfloat[mode~I]{\includegraphics[height=4.0cm]{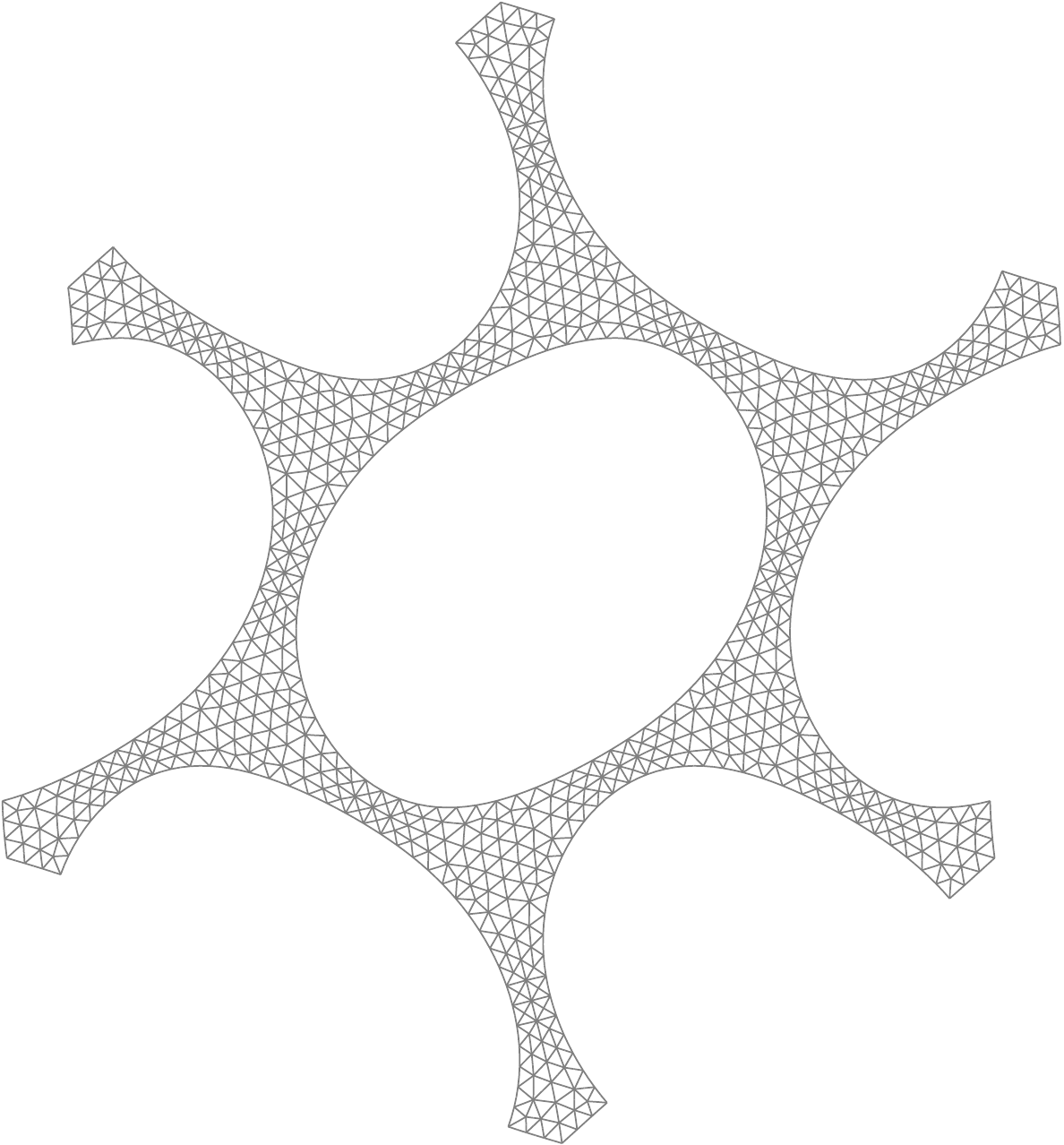}\label{Figure.sketch_3b}}\hspace{0.25em}
      \subfloat[mode~II]{\includegraphics[height=4.0cm]{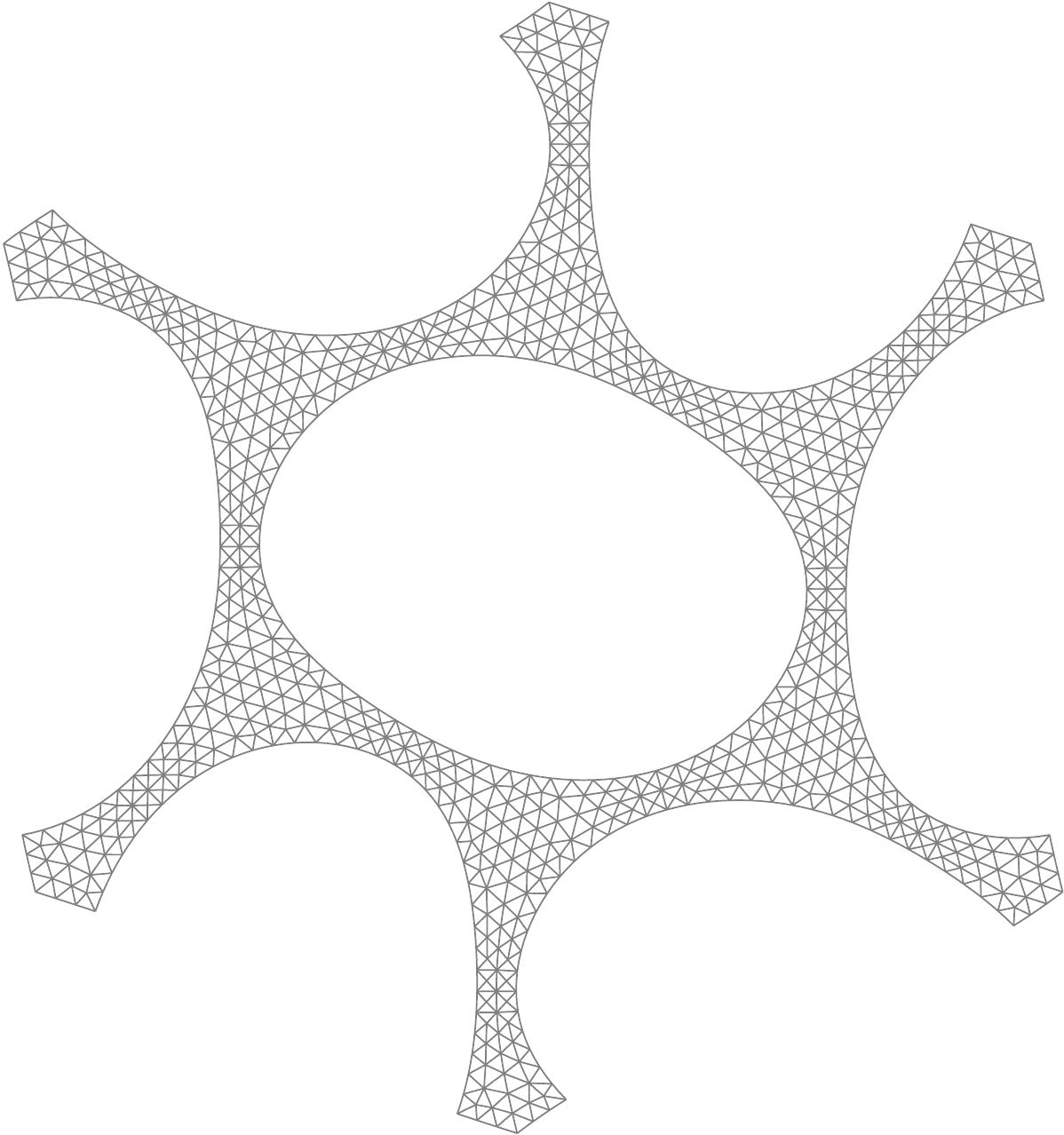}\label{Figure.sketch_3c}}\hspace{0.25em}
      \subfloat[mode~III]{\includegraphics[height=4.0cm]{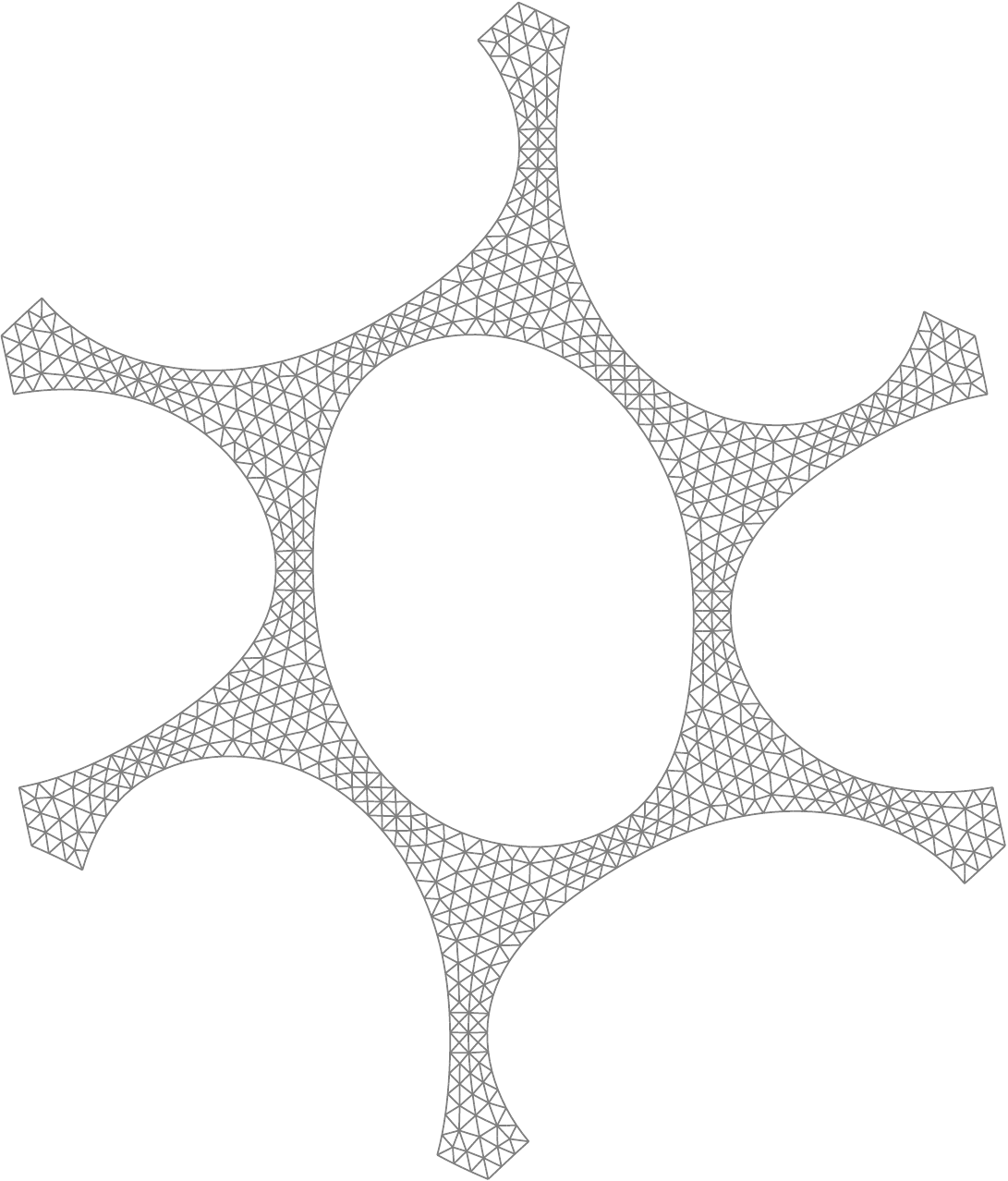}\label{Figure.sketch_3d}}
      \caption{Adopted periodic cell~$Q$ in~(a), along with three fundamental shear modes perpendicular to the three directions of cell walls. (b)~Mode~I, $\vec{\varphi}_1$, perpendicular to~$\theta = 90^\circ$, (c)~mode~II, $\vec{\varphi}_2$, perpendicular to~$\theta = 30^\circ$, and~(c) mode~III, $\vec{\varphi}_3$, perpendicular to~$\theta = - 30^\circ$.}
      \label{problem:fig3}
\end{figure}

At the triple bifurcation point, occurring for the equi-biaxial case~$\gamma = 1$, any linear combination of observed patterns can in principle be considered on the tangent space as potential deformation. \cite{Ohno2002} and~\cite{Okumura2002} defined a set of additional kinematic constraints on the honeycomb microstructure that results in a unique definition of mutually orthogonal patterns with a clear physical interpretation. These patterns are called \emph{modes} in what follows, and are denoted~$\vec{\varphi}_i(\vec{X})$, $i = 1, 2, 3$. These modes correspond to the shear mode (or pattern~I) developing perpendicular to each of the cell wall directions, i.e. at~$\theta = 90^\circ$ and~$\pm 30^\circ$. Three fundamental modes are thus defined. For the purpose of this contribution, the following analytic approximation is adopted for the first mode
\begin{equation}
\vec{\varphi}_1(\vec{X}) = \frac{1}{C} \left[ \sin\left(\frac{2\pi X_2}{\sqrt{3}\ell}\right)\vec{e}_1 + \frac{1}{\sqrt{3}}\sin\left(\frac{2 \pi X_1}{\ell}\right)\vec{e}_2 \right],
\label{eq:mode_1}
\end{equation}
were~$C$ is a normalization constant ensuring that~$\frac{1}{|Q|}\int_Q\|\vec{\varphi}_1(\vec{X})\|_2\,\mathrm{d}\vec{X} = 1$, where~$\|\vec{a}\|_2$ denotes the Euclidean norm of a vector~$\vec{a}$. The two remaining modes~II and~III are obtained by rotating~$\vec{\varphi}_1$ by~$\mp60^\circ$. The corresponding deformed configurations of the unit cell are shown in Figs.~\ref{Figure.sketch_3c}--\ref{Figure.sketch_3d}.

The individual modes, $\vec{\varphi}_i$, can be linearly combined to obtain the earlier defined patterns, $\vec{\pi}_i$, as follows
\begin{equation}
\begin{aligned}
\vec{\pi}_1 &= \vec{\varphi}_1, \\
\vec{\pi}_2 &= \vec{\varphi}_2 + \vec{\varphi}_3, \\
\vec{\pi}_3 &= \vec{\varphi}_1 + \vec{\varphi}_2 + \vec{\varphi}_3.
\end{aligned}
\label{eq:mode_combinations}  
\end{equation}
Note that various other linear combinations of modes exist resulting in various spatially shifted versions of the three patterns. For instance, if the signs of all the three modes are reversed for pattern~III, the directions of the petals (corresponding to the flower-like pattern) are reversed. If the signs of one or two of the modes are reversed, then the position of the undeformed (purely rotated) hole is shifted in space. Analogous behaviour applies also to the other patterns. In what follows, we will employ the \emph{modes}~$\vec{\varphi}_i$ to characterize the patterning which may occur in the microstructure. Patterns I--III should be the \emph{outcome} of the analysis for the appropriate loading.
%
%
\section{Extended Micromorphic Homogenization}
\label{homogenization}
%
%
\subsection{Kinematic Decomposition}
\label{sect:kinematic_decomposition}
The extended micromorphic computational homogenization scheme is based on a decomposition of the kinematic field~$\vec{u}(\vec{X})$, assumed to be of the form
\begin{equation}
\vec{u}(\vec{X}) = \vec{v}_0(\vec{X}) + \sum_{i=1}^{n} v_i(\vec{X})\vec{\varphi}_i(\vec{X}) + \vec{w}(\vec{X}),
\label{problem:eq5}
\end{equation}
which is a generalization of the ansatz of~\cite{Rokos2019}. The procedure outlined below follows this earlier work, and hence we summarize it here only briefly, referring to~\cite{Rokos2019} for more details. In Eq.~\eqref{problem:eq5}, $\vec{\varphi}_i(\vec{X})$, $i = 1,\dots, n$, are periodic deformation modes which are characteristic of the microstructure considered, establishing non-local interaction emerging from kinematic compatibility between neighbouring buckled periodic cells. They can be obtained in multiple ways: (i)~estimated analytically from full-scale DNS simulations, see~\cite{Rokos2019}, and Section~\ref{RVE_modes}, Eq.~\eqref{eq:mode_1}, above; (ii)~using a Bloch-type analysis, see e.g.~\cite{Bertoldi2008d}, which considers all possible directions and wave-lengths of buckling modes; (iii)~using a linearised buckling analysis, considering only wave-lengths that can be accommodated within a single RVE; or~(iv) identified experimentally through observed patterning modes of manufactured specimens emerging in the microstructure upon compression, see e.g.~\cite{Siavash:2020}. For the case of hexagonally stacked voided microstructures, the three modes~$\vec{\varphi}_i(\vec{X})$, $i = 1,\dots,3$, as defined in Eqs.~\eqref{eq:mode_1}--\eqref{eq:mode_combinations} are used, cf. also Fig.~\ref{problem:fig3}. The individual modes are periodic over~$Q$, with zero mean, whereas their amplitudes at the macroscale are governed by the scalar fields~$v_i(\vec{X})$, acting as amplitude modulators. The modes are expected to be active when the corresponding loading triggers an instability in the cell walls, capturing thereby the long-range fluctuations of the microstructure. Multiple modes can be activated through non-zero~$v_i(\vec{X})$ fields at the same time, but they are expected to be inactive until bifurcation occurs (i.e., pre-bifurcation, $v_i(\vec{X}) = 0$). When inappropriate deformation modes~$\vec{\varphi}_i$ are chosen by coincidence, or if one of the significant modes is neglected, the effect of those modes on the non-local behaviour of the microstructure is lost. This in turn induces errors which decrease proportionally with increasing scale ratio, eventually vanishing in the limit of infinite separation of scales. Furthermore, $\vec{v}_0(\vec{X})$ corresponds to the mean of the total displacement field~$\vec{u}(\vec{X})$, whereas the final term, $\vec{w}(\vec{X})$, is the conventional local microfluctuation field. This component addresses all kinematics not incorporated in the mean and the individual modes~$\vec{\varphi}_i(\vec{X})$ in the post-bifurcation regime, as well as all fast fluctuations in the pre-bifurcation regime.

Following the procedure of~\cite[][Section~5]{Rokos2019}, all effective fields~$\vec{v}_0$ and~$v_i$ are assumed to vary slowly compared to~$\vec{w}$ and~$\vec{\varphi}_i$, meaning that they can be approximated in a small neighbourhood around each point of interest~$\vec{X}$ spanned by a microscopic coordinate~$\vec{X}_\mathrm{m} \in \Omega_\mathrm{m}$ through a first order Taylor expansion as
\begin{equation}
\vec{u}(\vec{X},\vec{X}_\mathrm{m}) \approx \vec{v}_0(\vec{X}) + \vec{X}_\mathrm{m}\cdot\vec{\nabla}\vec{v}_0(\vec{X}) 
+ \sum_{i=1}^{n}
[v_i(\vec{X}) + \vec{X}_\mathrm{m}\cdot\vec{\nabla}v_i(\vec{X})]\vec{\varphi}_i(\vec{X}_\mathrm{m}) 
+
\vec{w}(\vec{X},\vec{X}_\mathrm{m}).
\label{fe2:eq2}
\end{equation}
In what follows, the close neighbourhood~$\Omega_\mathrm{m}$ of~$\vec{X}$ coincides with the periodic cell~$Q$ introduced in Section~\ref{micro_prop}, centred on~$\vec{X}$ and it will be referred to as a Representative Volume Element~(RVE) in order to establish a direct connection to the conventional computational homogenization framework and to distinguish it from the property of the microstructure~$Q$.

Upon taking into account the approximation of Eq.~\eqref{fe2:eq2}, uniqueness of the decomposition is ensured through an additional set of orthogonality conditions imposed on~$\vec{w}(\vec{X},\vec{X}_\mathrm{m})$, namely
\begin{equation}
\begin{aligned}
\int_{\Omega_\mathrm{m}} \vec{w}(\vec{X},\vec{X}_\mathrm{m}) \,\mathrm{d}\vec{X}_\mathrm{m} &= \vec{0}, \\
\int_{\Omega_\mathrm{m}} \vec{w}(\vec{X},\vec{X}_\mathrm{m}) \cdot \vec{\varphi}_i(\vec{X}_\mathrm{m})\,\mathrm{d}\vec{X}_\mathrm{m} &= 0, \quad i=1,\dots,n,\\
\int_{\Omega_\mathrm{m}} \vec{w}(\vec{X},\vec{X}_\mathrm{m}) \cdot [ \vec{\varphi}_i(\vec{X}_\mathrm{m}) \vec{X}_\mathrm{m} ]\,\mathrm{d}\vec{X}_\mathrm{m} &= \vec{0}, \quad i=1,\dots,n,\\
\end{aligned}
\label{eq:constraints}
\end{equation}
where the first condition ensures zero average of~$\vec{w}$ over~$\Omega_\mathrm{m}$, whereas the latter two conditions ensure orthogonality of~$\vec{w}$ with respect to~$\vec{\varphi}_i$ and~$\vec{\varphi}_i\vec{X}_\mathrm{m}$. In addition to the constraints of Eq.~\eqref{eq:constraints}, the microfluctuation field is assumed to be periodic over~$\Omega_\mathrm{m}$, i.e.
\begin{equation}
\llbracket \vec{w}(\vec{X},\vec{X}_\mathrm{m}) \rrbracket = \vec{0}, \quad \vec{X}_\mathrm{m} \in \partial\Omega_\mathrm{m}^+,
\label{eq:periodicity}
\end{equation}
where the RVE boundary~$\partial\Omega_\mathrm{m} = \partial\Omega_\mathrm{m}^+ \cup \partial\Omega_\mathrm{m}^-$ is split into two image and mirror parts, and~$\llbracket \vec{w}(\vec{X},\vec{X}_\mathrm{m}) \rrbracket = \vec{w}(\vec{X},\partial\Omega_\mathrm{m}^+) - \vec{w}(\vec{X},\partial\Omega_\mathrm{m}^-)$ denotes the jump of the field~$\vec{w}(\vec{X},\vec{X}_\mathrm{m})$ on the RVE boundary. Note that, following the same argumentation as typically used in first-order homogenization, orthogonality to the second term on the right-hand-side of Eq.~\eqref{fe2:eq2} (i.e.~orthogonality to~$\vec{X}_\mathrm{m}\cdot\vec{\nabla}\vec{v}_0(\vec{X})$, or~$\int_{\Omega_\mathrm{m}}\vec{w}(\vec{X},\vec{X}_\mathrm{m})\vec{X}_\mathrm{m}\,\mathrm{d}\vec{X}_\mathrm{m} = \bs{0}$ for arbitrary~$\vec{\nabla}\vec{v}_0$) is not enforced explicitly, but is ensured by the periodicity constraint for~$\vec{w}$ in Eq.~\eqref{eq:periodicity}.
%
%
\subsection{Definition of Potential Energy}
\label{sect:energy}
All constraints of Eqs.~\eqref{eq:constraints} and~\eqref{eq:periodicity} are enforced through Lagrange multipliers~$\vec{\mu}$, $\nu_i$, $\vec{\eta}_i$, and~$\vec{\lambda}_i$, and hence all unknown fields, i.e.~$\vec{v}_0(\vec{X})$, $v_i(\vec{X})$, and~$\vec{w}(\vec{X},\vec{X}_\mathrm{m})$, are obtained by minimizing the extended potential energy (i.e.~the Lagrangian~$\mathcal{L} = \mathcal{E}+\mathcal{C}$),
\begin{equation}
( \vec{v}_0,v_1,\dots,v_n,\vec{w} )
\in
\mathrm{arg}\ \underset{\vec{v}_0,v_1,\dots,v_n,\vec{w}}{\mbox{min}} \ \underset{\scriptsize
\vec{\mu},\ \underline{\nu},\ \underline{\eta},\ \vec{\lambda}}{\mbox{max}} \ \underbrace{{\mathcal{E}(\vec{u}) + \mathcal{C}(\vec{u},\vec{\mu},\underline{\nu},\underline{\eta},\vec{\lambda})}}_{\mathcal{L}(\vec{u},\vec{\mu},\underline{\nu},\underline{\eta},\vec{\lambda})},
\label{eq:energy}	
\end{equation}
where the displacement field~$\vec{u}$ is a function of all macroscopic fields~$\vec{v}_0,v_1,\dots,v_n$, and the microfluctuation field~$\vec{w}$, and where we have introduced~$\underline{\nu}(\vec{X}) = [\nu_1(\vec{X}), \dots, \nu_n(\vec{X})]$, $\underline{\eta}(\vec{X}) = [\vec{\eta}_1(\vec{X}), \dots, \vec{\eta}_n(\vec{X})]$ for brevity---see below for the definition of these quantities. The total potential energy~$\mathcal{E}$ and constraint term~$\mathcal{C}$ together define the Lagrangian~$\mathcal{L}$, with individual terms expressed as
\begin{equation}
\mathcal{E}(\vec{u}(\vec{X},\vec{X}_\mathrm{m})) = \frac{1}{|\Omega_\mathrm{m}|}\int_{\Omega} \int_{\Omega_\mathrm{m}} \Psi(\vec{X}_\mathrm{m}, \bs{F} (\vec{u}(\vec{X},\vec{X}_\mathrm{m}))) \,\mathrm{d}\vec{X}_\mathrm{m}\mathrm{d}\vec{X}, 
\label{eq:energy_definition}
\end{equation}
\begin{equation}
\begin{aligned}
\mathcal{C}(\vec{u}(\vec{X},\vec{X}_\mathrm{m}),\vec{\mu}(\vec{X})&,\underline{\nu}(\vec{X}),\underline{\eta}(\vec{X}),\vec{\lambda}(\vec{X},\vec{X}_\mathrm{m})) = \frac{1}{|\Omega_\mathrm{m}|}\int_{\Omega}
\bigg[
- \int_{\partial\Omega_\mathrm{m}^+}\vec{\lambda}(\vec{X},\vec{X}_\mathrm{m})\cdot \llbracket \vec{w}(\vec{X},\vec{X}_\mathrm{m}) \rrbracket \,\mathrm{d}\vec{X}_\mathrm{m} \\
&+ \vec{\mu}(\vec{X}) \cdot \int_{\Omega_\mathrm{m}} \vec{w}(\vec{X},\vec{X}_\mathrm{m}) \,\mathrm{d}\vec{X}_\mathrm{m} 
+ \sum_{i=1}^{n} \nu_i(\vec{X}) \int_{\Omega_\mathrm{m}} \vec{w}(\vec{X},\vec{X}_\mathrm{m}) \cdot \vec{\varphi}_i(\vec{X}_\mathrm{m})\,\mathrm{d}\vec{X}_\mathrm{m} \\
&+ \sum_{i=1}^{n} \vec{\eta}_i(\vec{X}) \cdot \int_{\Omega_\mathrm{m}} \vec{w}(\vec{X},\vec{X}_\mathrm{m}) \cdot [ \vec{\varphi}_i(\vec{X}_\mathrm{m}) \vec{X}_\mathrm{m} ]\,\mathrm{d}\vec{X}_\mathrm{m}
\bigg]\,\mathrm{d}\vec{X},
\end{aligned}	
\label{eq:constraint_definition}
\end{equation}
where~$\Psi(\vec{X}_\mathrm{m}, \bs{F} (\vec{u}(\vec{X},\vec{X}_\mathrm{m}))) = \chi(\vec{X}_\mathrm{m})\psi(\bs{F}(\vec{u}(\vec{X},\vec{X}_\mathrm{m})))$ is the elastic energy density of the microstructure, defined through a material indicator function~$\chi(\vec{X}_\mathrm{m})$ and Eq.~\eqref{problem:eq1}. The elastic energy density~$\Psi$ is assumed to be identical for all macroscopic points~$\vec{X}$, i.e.~for all RVEs, effectively resulting in a homogeneous microstructure in accordance to Section~\ref{problem}. The Lagrange multipliers in the constraint term~$\mathcal{C}$ are considered as constants inside each RVE for the bulk constraints of Eq.~\eqref{eq:constraints}, i.e.~$\vec{\mu}(\vec{X})$, $\underline{\nu}(\vec{X})$, and~$\underline{\eta}(\vec{X})$, whereas the Lagrange multipliers for the periodicity constraints are continuous functions on the boundary of each RVE, i.e.~$\vec{\lambda}(\vec{X},\vec{X}_\mathrm{m})$.
%
%
\subsection{Energy Minimization and Governing Equations}
\label{sect:governing_equations}
Taking the first variation of the Lagrangian~$\mathcal{L}$ and requiring it to vanish provides
\begin{equation}
\begin{aligned}
&0 = \delta \mathcal{L}(\vec{u},\vec{\mu},\underline{\nu},\underline{\eta},\vec{\lambda};\delta\vec{u},\delta\vec{\mu},\delta\underline{\nu},\delta\underline{\eta},\delta\vec{\lambda}) = \int_{\Omega}\bigg[
\int_{\Omega_\mathrm{m}}\bs{P}(\vec{X},\vec{X}_\mathrm{m}):\vec{\nabla}_\mathrm{m}\delta\vec{u}(\vec{X},\vec{X}_\mathrm{m})\,\mathrm{d}\vec{X}_\mathrm{m} \\
&- \int_{\partial\Omega_\mathrm{m}^+} \vec{\lambda}(\vec{X},\vec{X}_\mathrm{m})\cdot \llbracket \delta\vec{w}(\vec{X},\vec{X}_\mathrm{m}) \rrbracket + \delta\vec{\lambda}(\vec{X},\vec{X}_\mathrm{m})\cdot \llbracket \vec{w}(\vec{X},\vec{X}_\mathrm{m}) \rrbracket \,\mathrm{d}\vec{X}_\mathrm{m} \\
&+ \vec{\mu}(\vec{X}) \cdot \int_{\Omega_\mathrm{m}} \delta\vec{w}(\vec{X},\vec{X}_\mathrm{m}) \,\mathrm{d}\vec{X}_\mathrm{m} + \delta\vec{\mu}(\vec{X}) \cdot \int_{\Omega_\mathrm{m}} \vec{w}(\vec{X},\vec{X}_\mathrm{m}) \,\mathrm{d}\vec{X}_\mathrm{m} \\
&+ \sum_{i=1}^{n} \nu_i(\vec{X}) \int_{\Omega_\mathrm{m}} \delta\vec{w}(\vec{X},\vec{X}_\mathrm{m}) \cdot \vec{\varphi}_i(\vec{X}_\mathrm{m})\,\mathrm{d}\vec{X}_\mathrm{m} + \sum_{i=1}^{n}\delta\nu_i(\vec{X}) \int_{\Omega_\mathrm{m}} \vec{w}(\vec{X},\vec{X}_\mathrm{m}) \cdot \vec{\varphi}_i(\vec{X}_\mathrm{m})\,\mathrm{d}\vec{X}_\mathrm{m}\\
&+ \sum_{i=1}^{n} \vec{\eta}_i(\vec{X}) \cdot \int_{\Omega_\mathrm{m}} \delta\vec{w}(\vec{X},\vec{X}_\mathrm{m}) \cdot [ \vec{\varphi}_i(\vec{X}_\mathrm{m})\vec{X}_\mathrm{m} ]\,\mathrm{d}\vec{X}_\mathrm{m} \\
&+ \sum_{i=1}^{n} \delta\vec{\eta}_i(\vec{X})\cdot \int_{\Omega_\mathrm{m}} \vec{w}(\vec{X},\vec{X}_\mathrm{m}) \cdot [ \vec{\varphi}_i(\vec{X}_\mathrm{m})\vec{X}_\mathrm{m} ]\,\mathrm{d}\vec{X}_\mathrm{m}
\bigg]\,\mathrm{d}\vec{X},
\end{aligned}
\label{eq:firstvariation}
\end{equation}
where we have introduced the microscopic first Piola--Kirchhoff stress tensor, defined as
\begin{equation}
\bs{P}(\vec{X},\vec{X}_\mathrm{m}) = \frac{\partial\Psi(\vec{X}_\mathrm{m},\bs{F}(\vec{X},\vec{X}_\mathrm{m}))}{\partial\bs{F}^\mathsf{T}}.
\label{eq:stress}
\end{equation}
Taking the microscopic gradient~$\vec{\nabla}_\mathrm{m} = \partial/\partial X_{\mathrm{m},i}$ of~$\delta\vec{u}(\vec{X},\vec{X}_\mathrm{m})$ while considering the kinematic decomposition of Eq.~\eqref{fe2:eq2} yields
\begin{equation}
\begin{aligned}
\vec{\nabla}_\mathrm{m}\delta\vec{u}(\vec{X},\vec{X}_\mathrm{m}) &=  \vec{\nabla}\delta\vec{v}_0(\vec{X})
+ \sum_{i=1}^{n}\vec{\nabla}\delta v_i(\vec{X})\vec{\varphi}_i(\vec{X}_\mathrm{m}) \\
&+ \sum_{i=1}^{n} [\delta v_i(\vec{X}) + \vec{X}_\mathrm{m}\cdot\vec{\nabla} \delta v_i(\vec{X})]\vec{\nabla}_\mathrm{m}\vec{\varphi}_i(\vec{X}_\mathrm{m}) 
+ \vec{\nabla}_\mathrm{m}\delta\vec{w}(\vec{X},\vec{X}_\mathrm{m}),
\end{aligned}
\label{r_zeta_a}
\end{equation}
which, after substitution into Eq.~\eqref{eq:firstvariation}, making use of the divergence theorem, and assuming only essential boundary conditions are applied to~$\vec{v}_0$ on~$\Gamma_\mathrm{D} \subset \partial\Omega$, provides the following set of Euler--Lagrange equations
\begin{equation}
\delta\vec{v}_0:\ \left\{
\begin{aligned}
\vec{\nabla}\cdot\bs{\Theta}^\mathsf{T} &= \vec{0}, \quad \mbox{in}\ \Omega,\\
\bs{\Theta}\cdot\vec{N} &= \vec{0}, \quad \mbox{on}\ \Gamma_\mathrm{N}, \\
\end{aligned}\right.
\quad\delta v_i:\ \left\{
\begin{aligned}
\vec{\nabla}\cdot \vec{\Lambda}_{i}-\Pi_{i} &= 0, \quad \mbox{in}\ \Omega,\\
\vec{\Lambda}_{i}\cdot\vec{N} &= 0, \quad \mbox{on}\ \Gamma_\mathrm{N}, \\
\end{aligned}\right. \ i = 1, \dots, n,
\label{micromorphic:eq10}
\end{equation}
\begin{equation}
\begin{aligned}
\delta\vec{w}&:\ \left\{
\begin{aligned}
\vec{\nabla}_\mathrm{m}\cdot\bs{P}^\mathsf{T} &= \vec{\mu} + \sum_{i=1}^{n}\nu_i\vec{\varphi}_i + \sum_{i = 1}^{n}\vec{\eta}_i \cdot (\vec{\varphi}_i\vec{X}_\mathrm{m}), \ &&\mbox{in}\ \Omega_\mathrm{m}, \\
\bs{P}\cdot\vec{N}_\mathrm{m} &= \pm\vec{\lambda}, \ &&\mbox{on}\ \partial\Omega_\mathrm{m}^\pm,
\end{aligned}\right.\\
\delta\vec{\lambda}&:\ \mbox{periodicity constraint for~$\vec{w}$, Eq.~\eqref{eq:periodicity}},\\
\delta\vec{\mu}, \delta\underline{\nu}, \delta\underline{\eta}&:\ \mbox{kinematic constraints for~$\vec{w}$, Eq.~\eqref{eq:constraints}}.\\
\end{aligned}
\label{fe2:eq12}
\end{equation}
In Eqs.~\eqref{micromorphic:eq10} and~\eqref{fe2:eq12}, $\Gamma_\mathrm{N} \subset \partial\Omega$, $\Gamma_\mathrm{N} \cap \Gamma_\mathrm{D} = \emptyset$, denotes the part of the macroscopic boundary~$\partial\Omega$ along which zero tractions are applied, $\vec{N}$ is the corresponding macroscopic unit outer normal to~$\partial\Omega$, whereas~$\vec{N}_\mathrm{m}$ is RVE microscopic unit outer normal to~$\partial\Omega_\mathrm{m}$. The averaged macroscopic stress quantities, $\bs{\Theta}(\vec{X})$, $\Pi_{i}(\vec{X})$, and~$\vec{\Lambda}_{i}(\vec{X})$, have been introduced as
\begin{equation}
\begin{aligned}
\bs{\Theta}(\vec{X}) &= \frac{1}{|\Omega_\mathrm{m}|}\int_{\Omega_\mathrm{m}} \bs{P}(\vec{X},\vec{X}_\mathrm{m}) \, \mathrm{d}\vec{X}_\mathrm{m}, \\
\Pi_{i}(\vec{X}) &= \frac{1}{|\Omega_\mathrm{m}|}\int_{\Omega_\mathrm{m}} \bs{P}(\vec{X},\vec{X}_\mathrm{m}):\vec{\nabla}_\mathrm{m}\vec{\varphi}_i(\vec{X}_\mathrm{m}) \, \mathrm{d}\vec{X}_\mathrm{m}, \\
\vec{\Lambda}_{i}(\vec{X}) &= \frac{1}{|\Omega_\mathrm{m}|}\int_{\Omega_\mathrm{m}} \bs{P}^\mathsf{T}(\vec{X},\vec{X}_\mathrm{m})\cdot\vec{\varphi}_i(\vec{X}_\mathrm{m}) + \vec{X}_\mathrm{m}[\bs{P}(\vec{X},\vec{X}_\mathrm{m}):\vec{\nabla}_\mathrm{m}\vec{\varphi}_i(\vec{X}_\mathrm{m})] \, \mathrm{d}\vec{X}_\mathrm{m}.
\end{aligned}
\label{fe2:eq10}
\end{equation}
In Eq.~\eqref{fe2:eq10}, each macroscopic point~$\vec{X}$ has associated with it an RVE with the domain~$\Omega_\mathrm{m}$, over which the local microscopic stress~$\bs{P}(\vec{X},\vec{X}_\mathrm{m})$ is defined, recall Eq.~\eqref{eq:stress}. From Eq.~\eqref{micromorphic:eq10}, and through Eqs.~\eqref{fe2:eq10}, \eqref{eq:stress}, and the form of~$\vec{\nabla}\vec{u}$ similar to Eq.~\eqref{r_zeta_a} featuring in the deformation tensor~$\bs{F}$, it can furthermore be	observed that all the homogenized stresses depend on the gradients of~$\vec{v}_0$ and~$v_i$, and hence a micromorphic continuum emerges. 

\begin{figure}
	\centering
	\def\svgwidth{0.7\textwidth}
	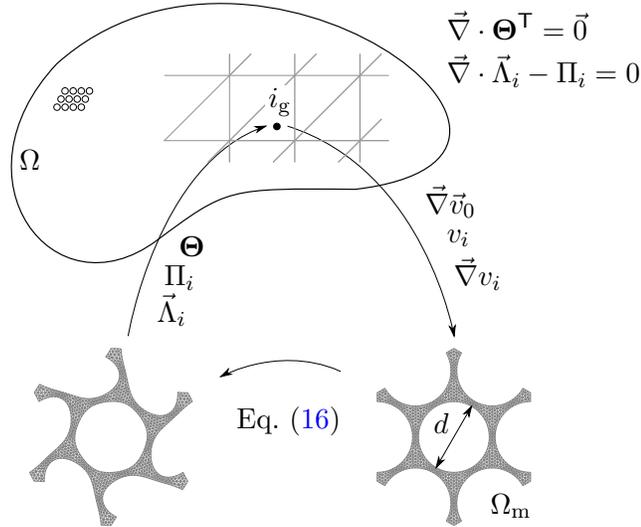
	\caption{Sketch of the micromorphic computational homogenization framework. At each macroscopic Gauss point~$i_\mathrm{g}$, macroscopic kinematic quantities~$\vec{\nabla}\vec{v}_0$, $v_i$, and~$\vec{\nabla}v_i$, $i = 1, \dots, n$, are prescribed to a periodic cell, or a Representative Volume Element~(RVE), with a domain~$\Omega_\mathrm{m}$, where the microfluctuation field~$\vec{w}$ is computed and condensed out. Homogenized properties (i.e.~stresses~$\bs{\Theta}$, $\Pi_i$, and~$\vec{\Lambda}_i$) are transferred back to the macroscale, where they are equilibrated.}
	\label{framework:fig6}
\end{figure}

The overall micromorphic computational homogenization procedure is summarized in Fig.~\ref{framework:fig6}, in which both the micro- and macro-boundary value problems are solved using the standard finite element method for the spatial discretization and a quasi-Newton algorithm for the energy minimization; see~\cite{Rokos2019} for further details. Inspecting Eqs.~\eqref{micromorphic:eq10}--\eqref{fe2:eq12} in combination with Fig.~\ref{framework:fig6}, it may become clear that the presented framework reduces to conventional computational homogenization as a special case if no patterning modes~$\vec{\varphi}_i$ are added in Eq.~\eqref{fe2:eq2}, or if all micromorphic fields vanish, i.e.~when~$ v_i = 0 $ and~$ \vec{\nabla}v_i = \vec{0} $, as observed e.g.~in the pre-bifurcation regime. In the case of large separation of scales, $ v_i \neq 0 $ while~$ \vec{\nabla}v_i = \vec{0}$. Because all~$ \vec{\varphi}_i $ as well as~$ \vec{w} $ are periodic, the micromorphic as well as conventional computational homogenization scheme provide exact solutions in such a situation, and the~$v_i$ fields merely correspond to the amount of the periodic patterning mode~$\vec{\varphi}_i$ present in the RVE deformation~$\vec{u}$. Only when~$ \vec{\nabla}v_i \neq \vec{0} $ holds, e.g.~due to applied boundary conditions or spatial mixing of patterning modes, resulting RVEs are non-periodic, strong non-local effects emerge, and micromorhpic solutions deviate from those of a (local) conventional computational homogenization scheme.
%
%
\section{Computational Assessment}
\label{results}
This section details results obtained using the micromorphic computational homogenization scheme of Section~\ref{homogenization} for the honeycomb microstructure introduced in Section~\ref{problem} (i.e.~for~$n = 3$) and various loading conditions. First, the evolution of a hexagonal honeycomb for two biaxiality ratios~$\gamma$ triggering temporal switching between individual patterns in a macroscopically infinite geometry is studied. Second, a finite cruciform-shaped specimen exhibiting spatially mixed patterns is presented.
%
%
\subsection{Temporal Switching of Patterns}
\label{temporal}
\begin{figure}[!t]
	\centering
	\subfloat[undeformed]{\includegraphics[height=4.3cm]{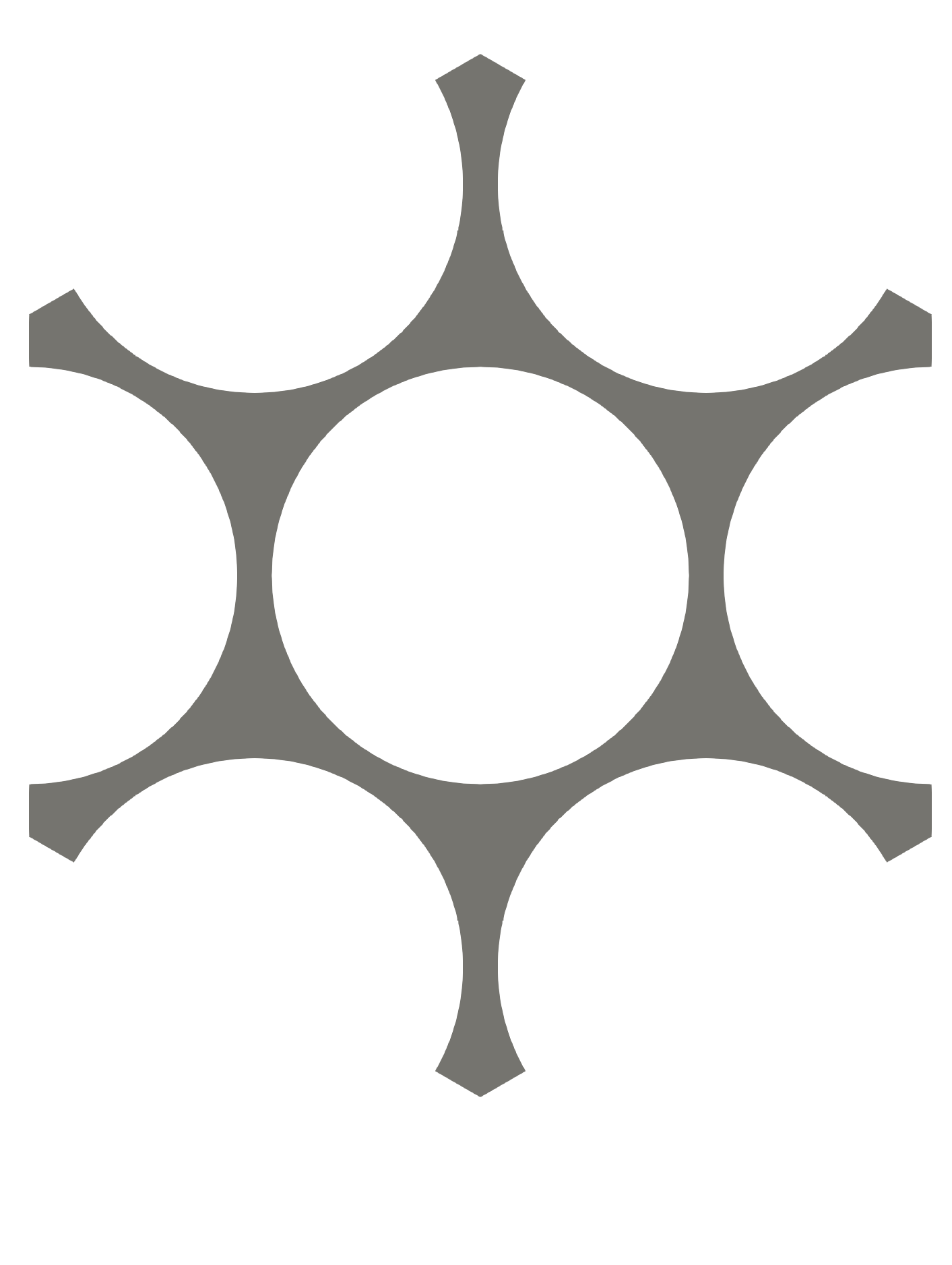}\label{temporal:I-IIIa}}
	\hspace{0.5em}
	\subfloat[initial pattern~I]{\includegraphics[height=4.3cm]{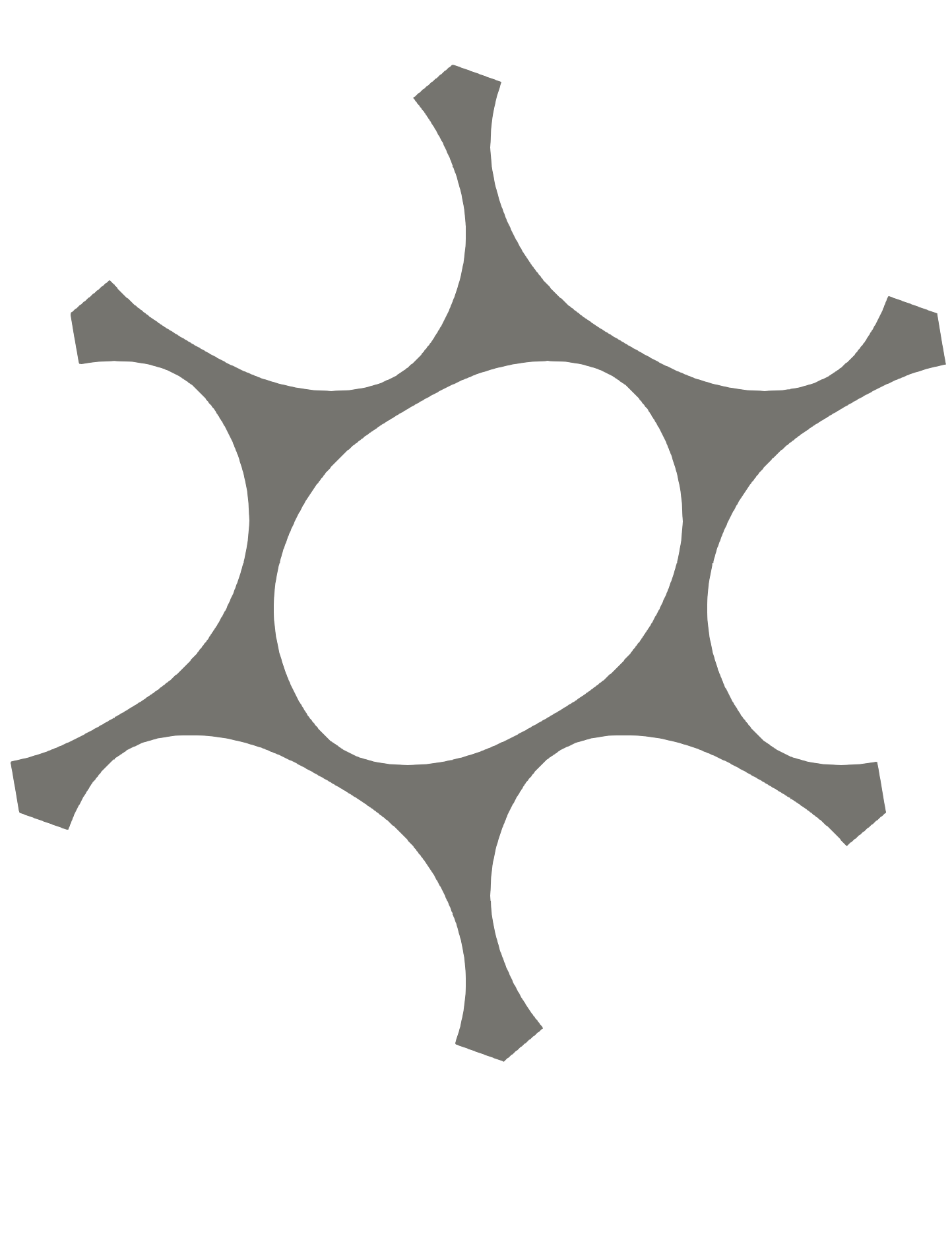}\label{temporal:I-IIIb}}
	\hspace{0.7em}
	\subfloat[final pattern~III$'$]{\includegraphics[height=4.3cm]{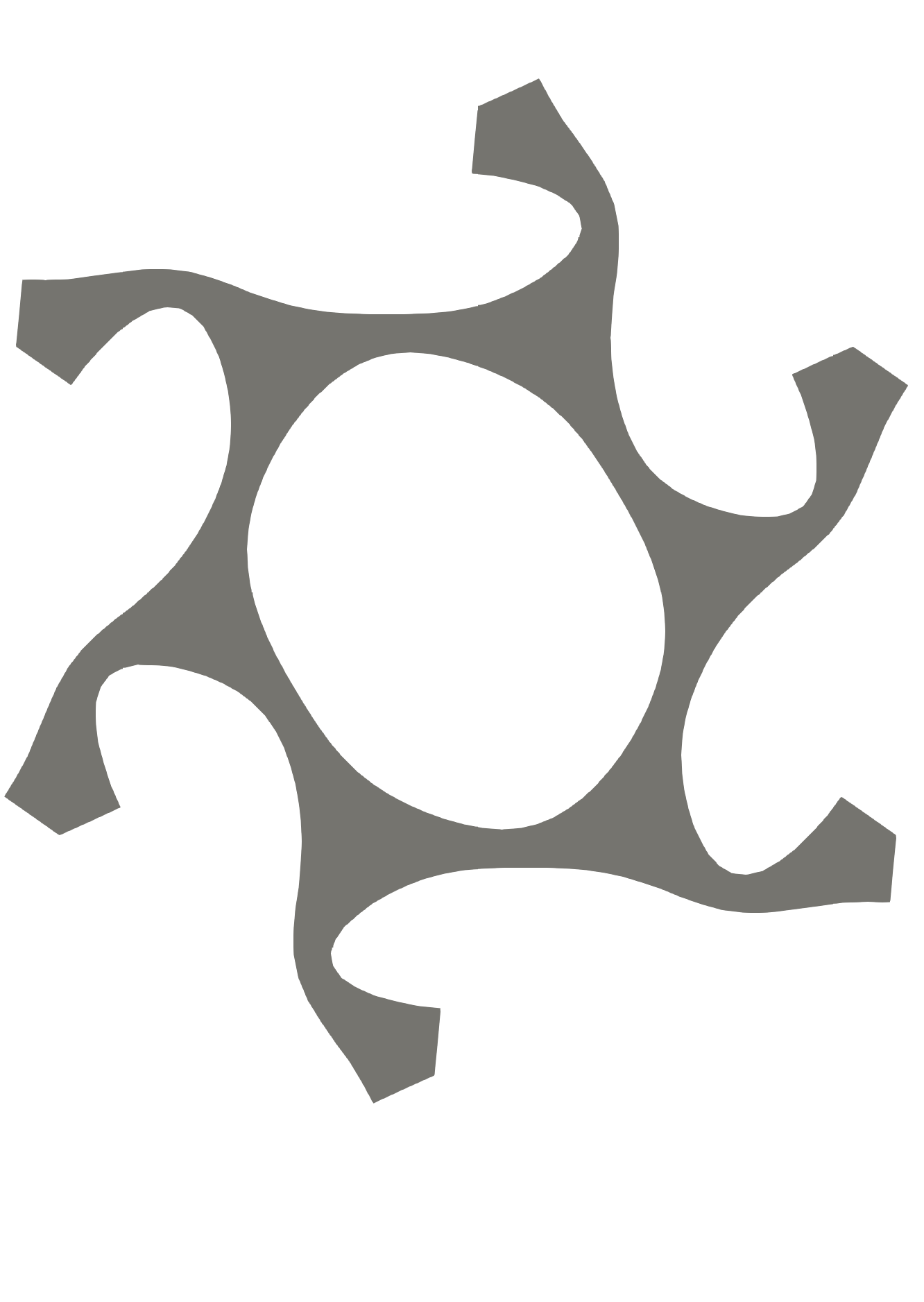}\label{temporal:I-IIIc}}
	\hfill
	\subfloat[stress--strain diagram]{\includegraphics[scale=0.9]{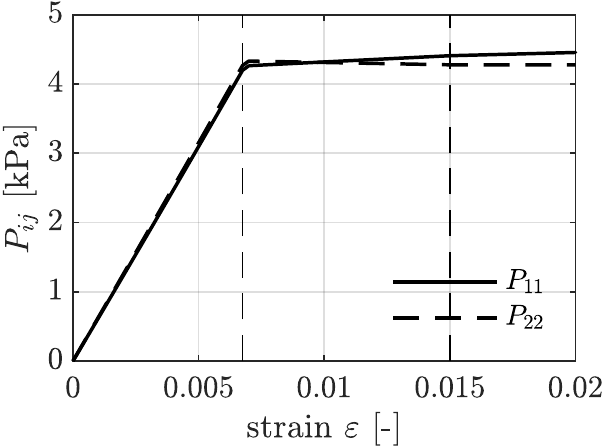}\label{temporal:I-IIId}}
	\caption{DNS solution for the I-III case of~$\gamma=1.15$. (a)~Undeformed configuration of an infinite circular honeycomb modelled with periodic boundary conditions. (b)~Deformed configuration for~$\varepsilon = 1\%$, showing pattern~I. (c)~Deformed configuration for~$\varepsilon = 5\%$, showing a distorted pattern~III$'$. The displacements of both deformed states are magnified by a factor of two. (d)~Corresponding nominal stress--strain diagram.}
	\label{temporal:I-III}
\end{figure}

\begin{figure}[!t]
	\centering
	\subfloat[undeformed]{\includegraphics[height=4.3cm]{time_Ia-crop.png}\label{temporal:II-IIIa}}
	\hspace{1em}
	\subfloat[initial pattern~II]{\includegraphics[height=4.3cm]{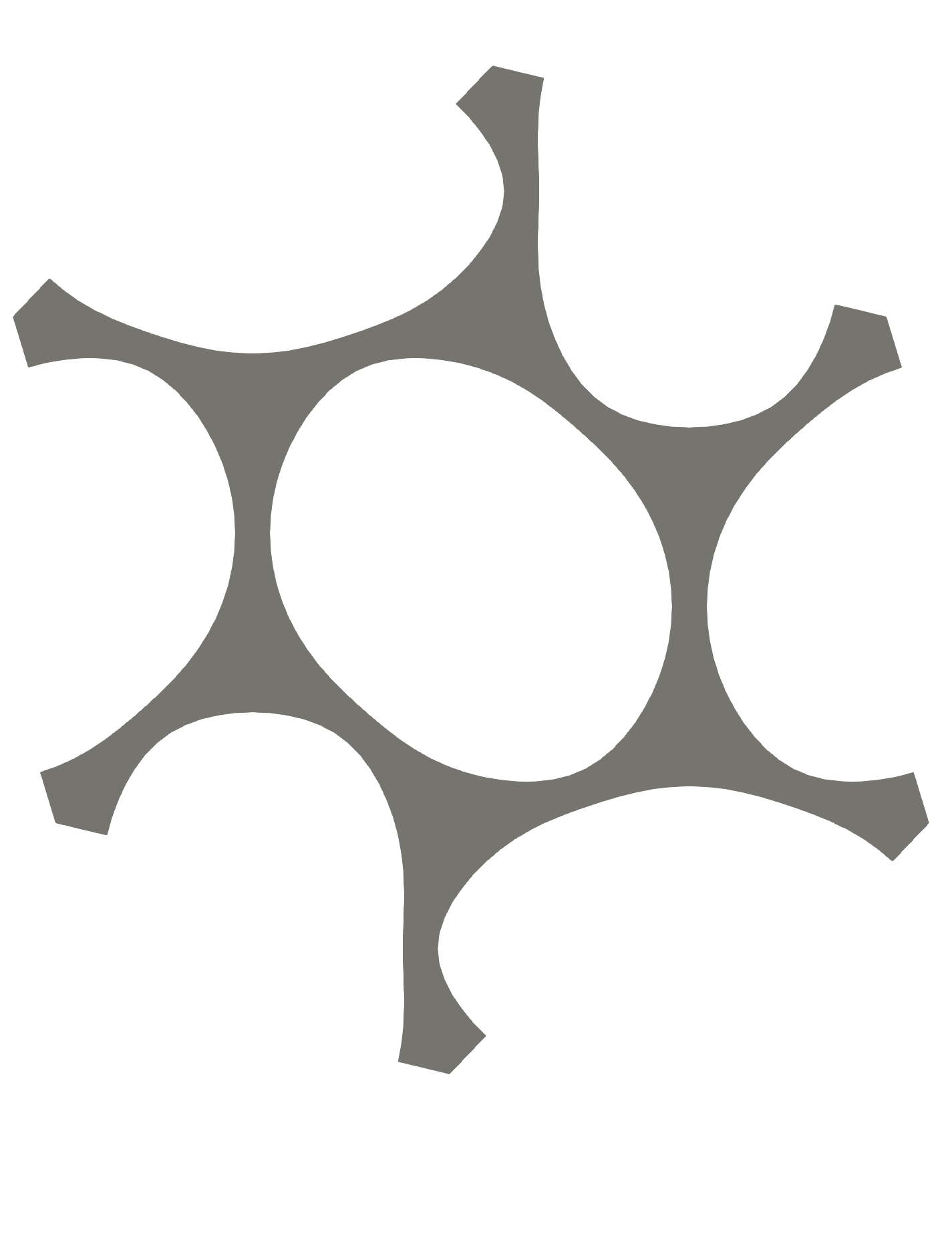}\label{temporal:II-IIIb}}
	\hspace{0.3em}
	\subfloat[final pattern~III$''$]{\includegraphics[height=4.3cm]{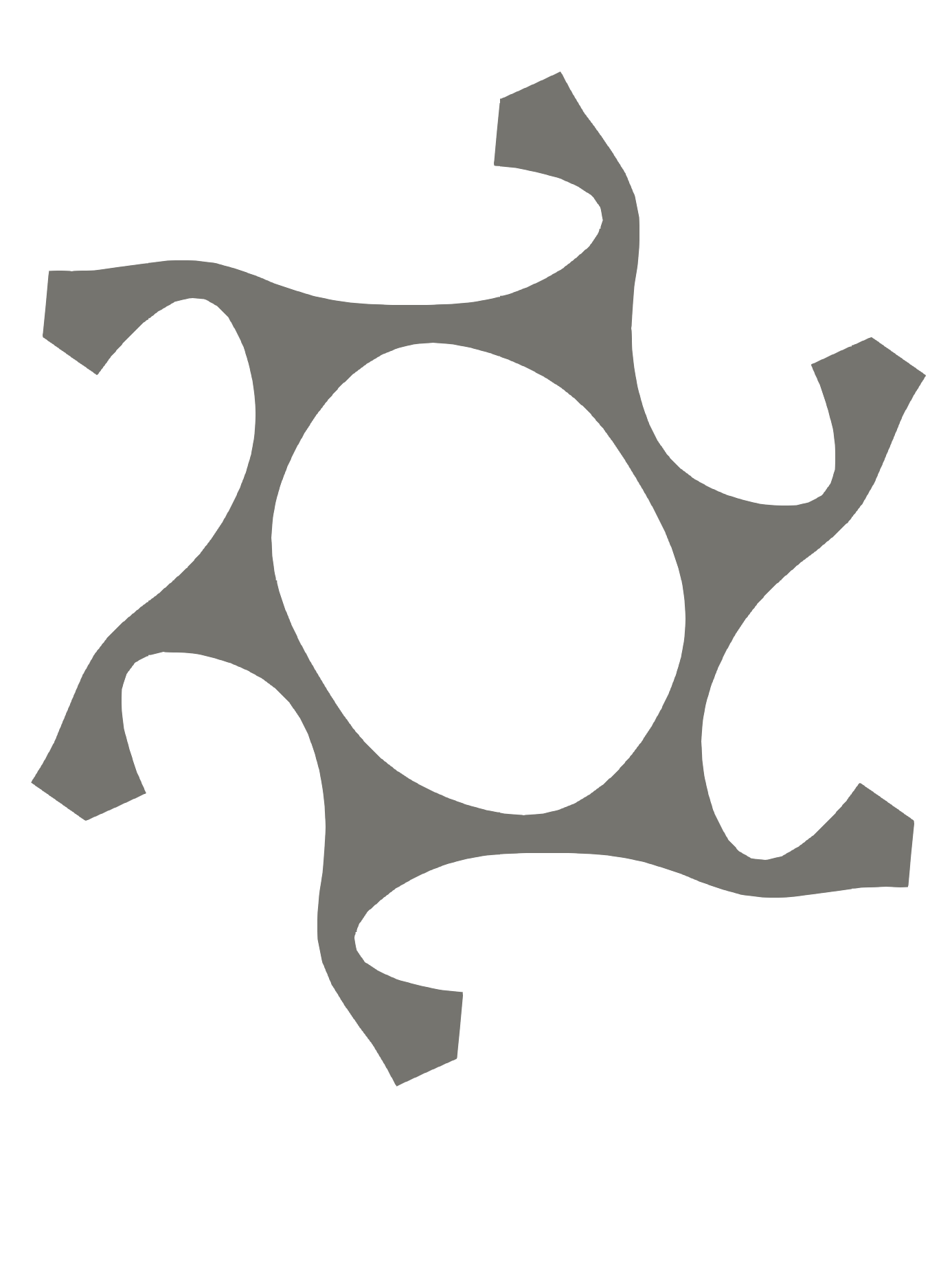}\label{temporal:II-IIIc}}
	\hfill
	\subfloat[stress--strain diagram]{\includegraphics[scale=0.9]{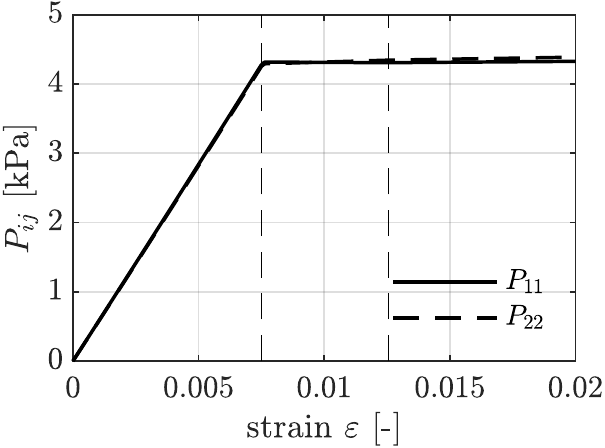}\label{temporal:II-IIId}}
	\caption{DNS solution for the II-III case of~$\gamma=0.95$. (a)~Undeformed configuration of an infinite circular honeycomb modelled with periodic boundary conditions. (b)~Deformed configuration for~$\varepsilon = 1\%$, showing pattern~II. (c)~Deformed configuration for~$\varepsilon = 5\%$, showing a distorted pattern~III$''$. The displacements of both deformed states are magnified by a factor of two. (d)~Corresponding nominal stress--strain diagram.}
	\label{temporal:II-III}
\end{figure}

\cite{Okumura2002} identified two load-cases, with biaxiality ratios~$\gamma=3/2$ and~$3/4$, for which infinite microstructures with a hexagonal stacking of hexagonal holes lead to two sequentially occurring bifurcation points. Here, we demonstrate this phenomenon for~$\gamma=1.15$ and~$\gamma=0.95$. Note that different biaxiality ratios~$\gamma$ are necessary because of the different honeycomb geometry considered here (circular holes instead of hexagonal ones). The DNS result corresponding to one RVE subjected to an overall deformation~$\bs{F}(\varepsilon) = \bs{I}+\bs{G}(\varepsilon)$, $\varepsilon \in [0, 5]\%$ being a scalar strain measure parametrizing the externally applied load, is shown in Figs.~\ref{temporal:I-III} and~\ref{temporal:II-III}, where the reference and two deformed configurations at different levels of overall applied strain are shown. For better clarity, displacements have been magnified by a factor of two. The corresponding nominal stress--strain diagrams are included as well, in which the two consecutive bifurcation points are marked by the vertical dashed lines. In Fig.~\ref{temporal:I-III}, i.e.~for the case of~$\gamma = 1.15$ and~$\bs{G}(\varepsilon) = -\varepsilon\,(\vec{e}_1\vec{e}_1+1.15\,\vec{e}_2\vec{e}_2)$, we observe that pattern~I emerges at the first bifurcation point, occurring at~$\varepsilon \approx 0.7\%$ (Fig.~\ref{temporal:I-IIIb}), whereas beyond the second bifurcation point, corresponding to~$\varepsilon \approx 1.5\%$, a distorted pattern~III$'$ develops (Fig.~\ref{temporal:I-IIIc}). This loading is therefore referred to as the I-III case. For the biaxiality ratio~$\gamma = 0.95$ and~$\bs{G}(\varepsilon) = -\varepsilon\,(\vec{e}_1\vec{e}_1+0.95\,\vec{e}_2\vec{e}_2)$, pattern~II emerges at the first bifurcation point, corresponding to~$\varepsilon \approx 0.8\%$ (Fig.~\ref{temporal:II-IIIb}, which is shifted as compared to Fig.~\ref{problem:fig2c}), followed by a distorted pattern~III$''$ after the second bifurcation point at~$\varepsilon \approx 1.3\%$ (Fig.~\ref{temporal:II-IIIc}), which resembles the one in Fig.~\ref{temporal:I-IIIc}. This loading is referred to as the II-III case.

\begin{figure}[!t]
	\centering
	\begin{tabular}{ccc}
		\subfloat[deformed RVE]{\includegraphics[height=4.3cm]{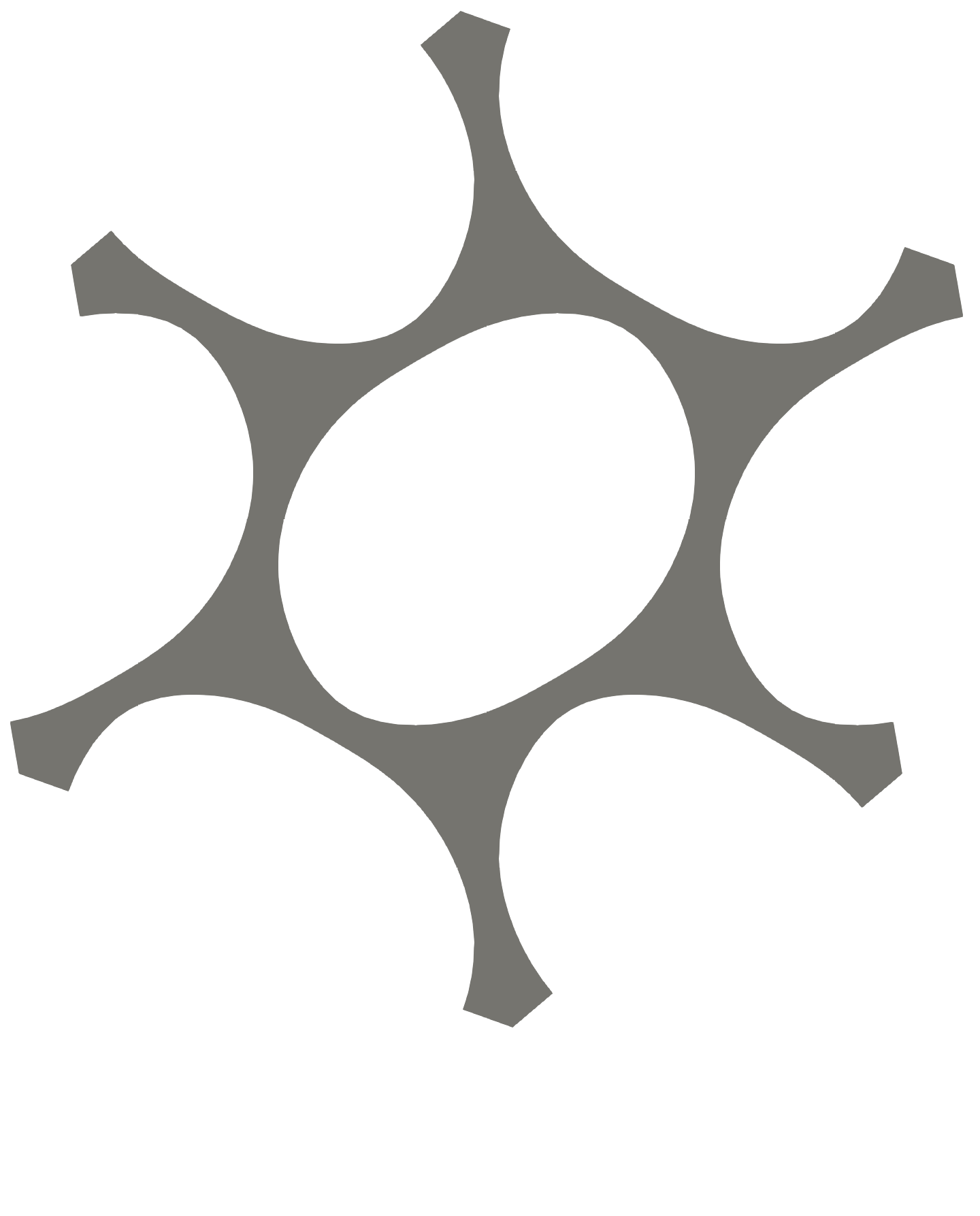}\label{micromorphic:I-IIIa}} &
		\subfloat[$\vec{e}_1$ component of~$\vec{v}_0$~\mbox{[mm]}]{\includegraphics[scale=0.9]{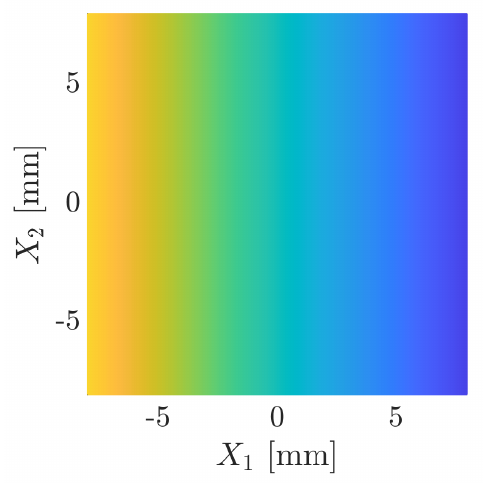}\label{micromorphic:I-IIIb}}&
		\subfloat[$\vec{e}_2$ component of~$\vec{v}_0$~\mbox{[mm]}]{\includegraphics[scale=0.9]{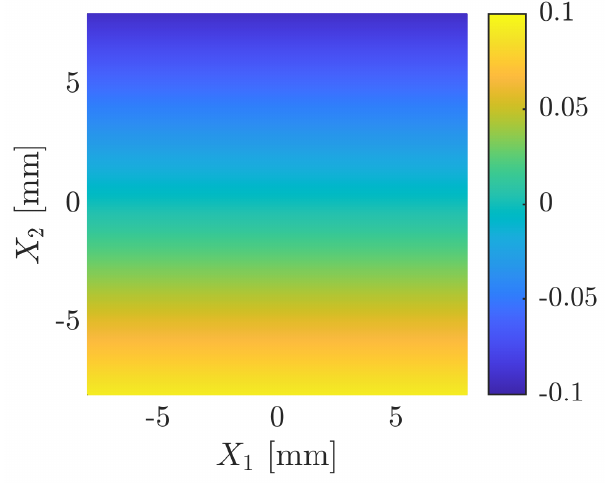}\label{micromorphic:I-IIIc}}\\ 
		\subfloat[$v_1$~\mbox{[mm]}]{\includegraphics[scale=0.9]{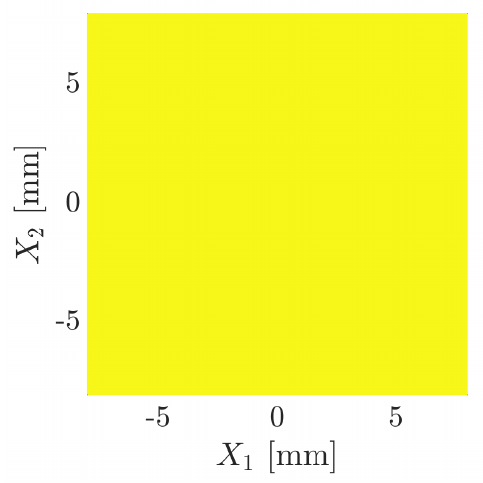}\label{micromorphic:I-IIId}}&
		\subfloat[$v_2$~\mbox{[mm]}]{\includegraphics[scale=0.9]{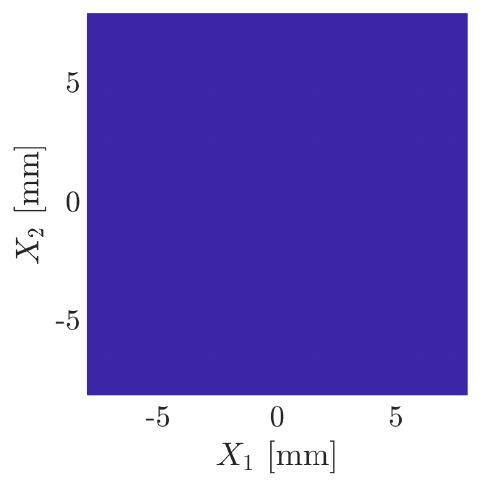}\label{micromorphic:I-IIIe}}&
		\subfloat[$v_3$~\mbox{[mm]}]{\includegraphics[scale=0.9]{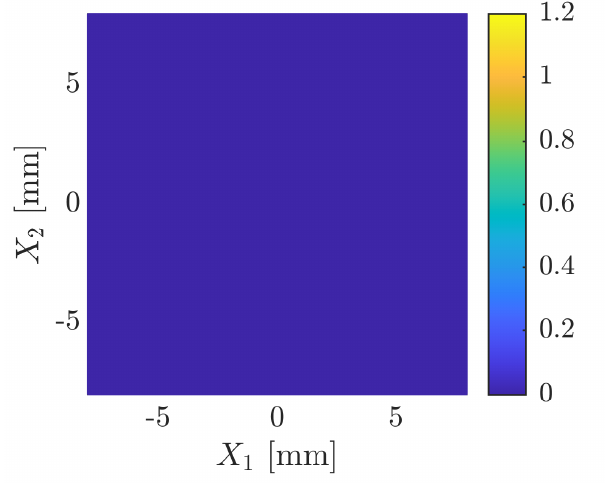}\label{micromorphic:I-IIIf}}
	\end{tabular}
	\caption{Micromorphic homogenized solution of temporal switching of patterns exhibited in the I-III case for biaxial compression~$\gamma=1.15$ at~$\varepsilon = 1\%$. (a)~Shape of the deformed RVE (deformation magnified by a factor of two). The two components of the macroscopic displacement field~$\vec{v}_0$ are shown in~(b) horizontal and~(c) vertical direction, whereas the spatial distributions of the individual micromorphic fields corresponding to the amplitudes of the modes~$\vec{\varphi}_i$ are shown in~(d)--(f).}
	\label{micromorphic:I-III}
\end{figure}
\begin{figure}[!t]
	\centering
	\begin{tabular}{ccc}	
		\subfloat[deformed RVE]{\includegraphics[height=4.3cm]{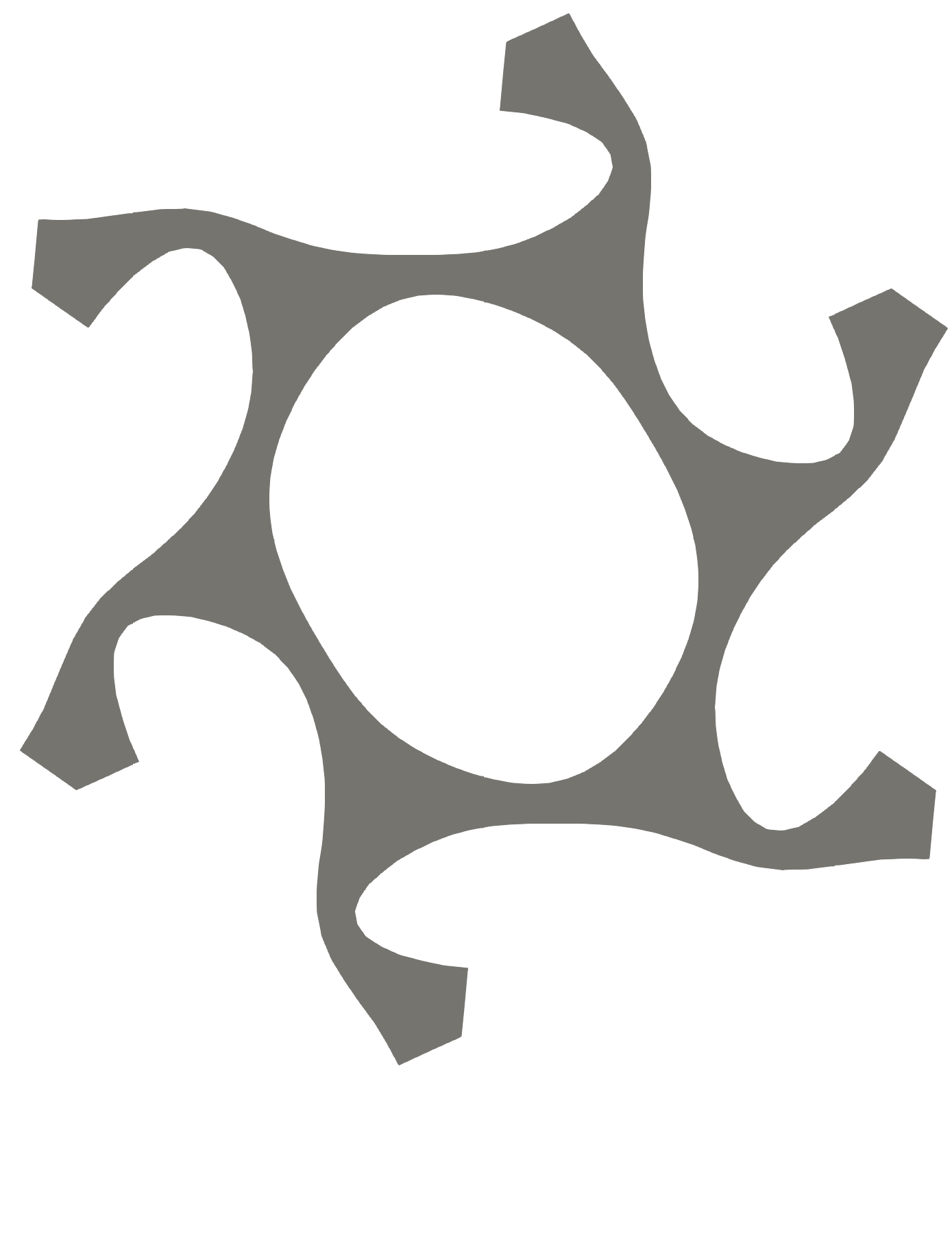}\label{micromorphic:II-IIIa}}&
		\subfloat[$\vec{e}_1$ component of~$\vec{v}_0$~\mbox{[mm]}]{\includegraphics[scale=0.9]{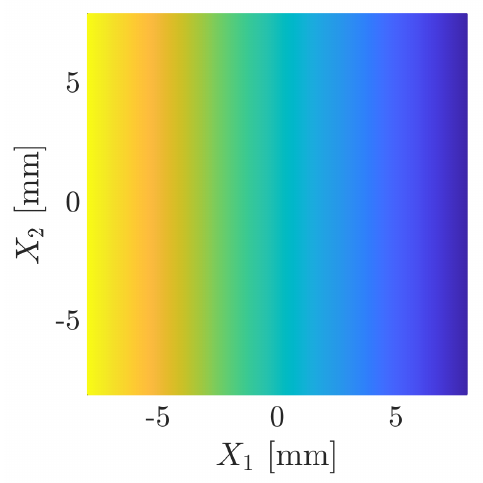}\label{micromorphic:II-IIIb}}&
		\subfloat[$\vec{e}_2$ component of~$\vec{v}_0$~\mbox{[mm]}]{\includegraphics[scale=0.9]{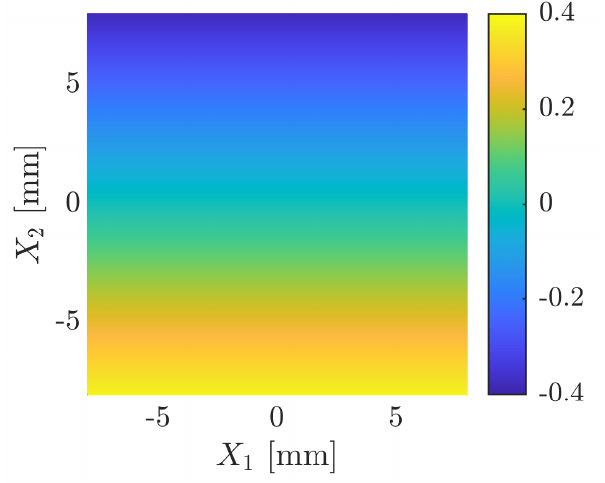}\label{micromorphic:II-IIIc}}\\ 
		\subfloat[$v_1$~\mbox{[mm]}]{\includegraphics[scale=0.9]{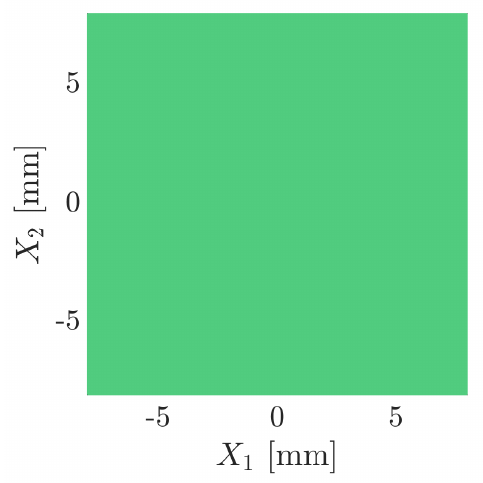}\label{micromorphic:II-IIId}}&
		\subfloat[$v_2$~\mbox{[mm]}]{\includegraphics[scale=0.9]{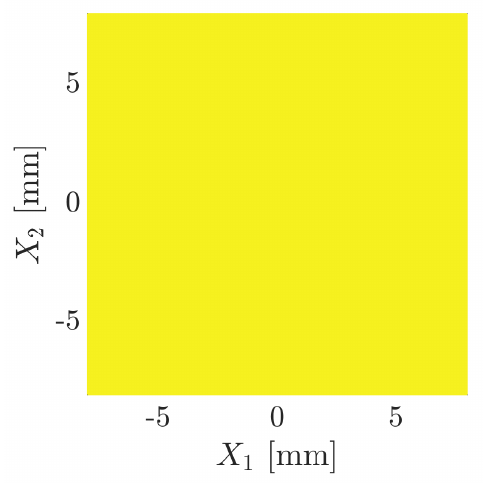}\label{micromorphic:II-IIIe}}&
		\subfloat[$v_3$~\mbox{[mm]}]{\includegraphics[scale=0.9]{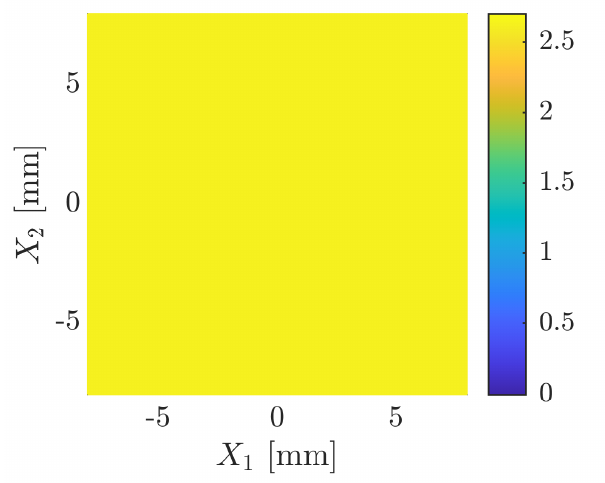}\label{micromorphic:II-IIIf}}
	\end{tabular}
	\caption{Micromorphic homogenized solution of temporal switching of patterns exhibited in the II-III case of biaxial compression~$\gamma=0.95$ at~$\varepsilon = 5\%$. (a)~Shape of the deformed RVE (deformation magnified by a factor of two). The two components of the macroscopic displacement field~$\vec{v}_0$ are shown in~(b) horizontal and~(c) vertical direction, whereas the spatial distributions of the individual micromorphic fields corresponding to the amplitudes of the modes~$\vec{\varphi}_i$ are shown in~(d)--(f).}
	\label{micromorphic:II-III}
\end{figure}

To assess the performance of the micromorphic computational homogenization scheme for switching of patterns in time, a macroscopically periodic square domain of size~$16 \times 16$~mm (approximately~$12 \times 12$ holes) is considered. This domain is discretized into eight equally-sized six-node quadratic triangular elements with the three-point Gauss integration rule, and subjected to the same overall prescribed deformation~$\bs{G}(\varepsilon)$ as the DNS solution, examined only for two snapshots corresponding to~$\varepsilon = 1\%$ and~$\varepsilon = 5\%$. The obtained results are presented in Fig.~\ref{micromorphic:I-III} for the I-III case and~$\varepsilon = 1\%$, and in Fig.~\ref{micromorphic:II-III} for the II-III case and~$\varepsilon = 5\%$. As expected, we clearly see in both figures that spatially linear displacement fields~$\vec{v}_0$ and constant mode amplitudes~$v_i$ result. The other two combinations are not shown for conciseness, but are fully consistent with these two.

For the I-III case we furthermore notice that the RVE's deformed shape (Fig.~\ref{micromorphic:I-IIIa}) matches well with the DNS result (Fig.~\ref{temporal:I-IIIb}), i.e.~it resembles pattern~I. This observation is supported by the resulting values of~$v_i$ presented in Figs.~\ref{micromorphic:I-IIId}--\ref{micromorphic:I-IIIf}, which attain~$v_1(\vec{X}) = 1.2$~mm, $v_2(\vec{X}) = 0.0$~mm, and~$v_3(\vec{X}) = 0.0$~mm, i.e.~only presence of mode~I (and hence pattern~I). For the second snapshot of~$\varepsilon = 5\%$ (not shown), the obtained displacement field~$\vec{v}_0$ is again linear, whereas~$v_1(\vec{X}) = 3.8$~mm, $v_2(\vec{X}) = 1.4$~mm, and~$v_3(\vec{X}) = 1.4$~mm, resembling a distorted pattern~III, as all three modes are active with significant magnitudes (recall the third row of Eq.~\eqref{eq:mode_combinations}). The corresponding nominal stresses~$\Theta_{22}(\varepsilon = 1\%) = -4.310$~kPa and~$\Theta_{22}(\varepsilon = 5\%) = -4.256$~kPa match exactly with the DNS predictions, see also Fig.~\ref{temporal:I-IIId}.

The II-III case achieves similar accuracy. For the first snapshot at~$\varepsilon=1\%$ (not shown), a linear field~$\vec{v}_0$ results, with magnitudes of the micromorphic fields corresponding to~$v_1(\vec{X}) = 0.0$~mm, $v_2(\vec{X}) = 0.7$~mm, and~$v_3(\vec{X}) = 0.7$~mm, clearly resembling pattern~II (recall the second row of Eq.~\eqref{eq:mode_combinations}). For the second snapshot at~$\varepsilon = 5\%$, the results are presented in Fig.~\ref{micromorphic:II-III}. Here we again notice that the deformed RVE shape of Fig.~\ref{micromorphic:II-IIIa} coincides well with the DNS result of Fig.~\ref{temporal:II-IIIc}, which is reflected by the magnitudes of the corresponding micromorphic fields, $v_1(\vec{X}) = 1.6$~mm, $v_2(\vec{X}) = 2.6$~mm, and~$v_3(\vec{X}) = 2.6$~mm. Obtained homogenized stresses~$\Theta_{22}(\varepsilon = 1\%) = -4.317$~kPa and~$\Theta_{22}(\varepsilon = 5\%) = -4.576$~kPa, exactly correspond to DNS predictions, cf. Fig.~\ref{temporal:II-IIId}.

The homogenized solutions presented above were obtained by initializing~$\vec{v}_0$ to the exact linear deformation specified by~$\bs{G}(\varepsilon)$, whereas the individual micromorphic fields were initialized towards the expected patterns. To verify that the presented results do not correspond to local minima, other mode combinations representing different patterns were considered as initial guesses as well. It was observed that the solutions either lead to the correct result, or if a different pattern occurs, it corresponds to a higher elastic energy (hence being less favourable). A more systematic approach is desirable for proper identification of patterning modes (using bifurcation analysis), which will be reported separately. When one of the modes is neglected in Eqs.~\eqref{problem:eq5}--\eqref{fe2:eq2}, say~$\vec{\varphi}_3$, exact results in terms of RVE deformations and stress--strain relations are still obtained. The effect of the periodic component~$v_3\vec{\varphi}_3$ simply transfers to the periodic microfluctuation field~$\vec{w}$. This is possible only for the case of infinite separation of scales, for which~$\vec{\nabla}v_3 = \vec{0}$ holds and deformed RVEs are periodic. When~$\vec{\nabla}v_3 \neq \vec{0}$, however, non-periodic RVEs and a non-local continuum emerge, which induce errors once any of the essential modes is neglected.
%
%
\subsection{Spatial Mixing of Patterns}
\label{spatial_mixing}
\begin{figure}[!t]
	\centering
	\subfloat[reference configuration]{\includegraphics[height=5.0cm]{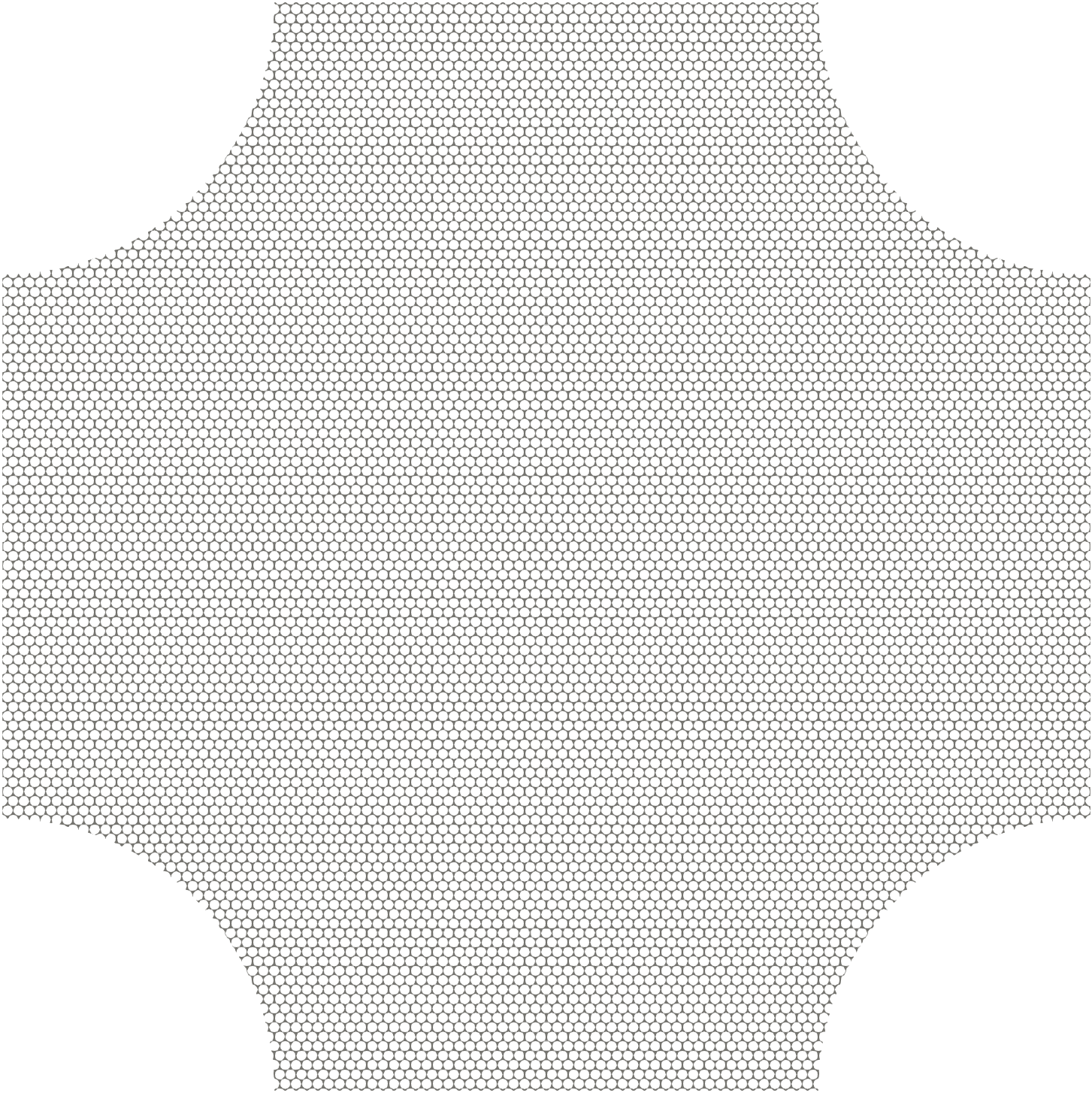}\label{cruciform}}
	\hspace{1.0em}
	\subfloat[deformed configuration]{\def\svgwidth{7.72cm}
\begingroup%
  \makeatletter%
  \providecommand\color[2][]{%
    \errmessage{(Inkscape) Color is used for the text in Inkscape, but the package 'color.sty' is not loaded}%
    \renewcommand\color[2][]{}%
  }%
  \providecommand\transparent[1]{%
    \errmessage{(Inkscape) Transparency is used (non-zero) for the text in Inkscape, but the package 'transparent.sty' is not loaded}%
    \renewcommand\transparent[1]{}%
  }%
  \providecommand\rotatebox[2]{#2}%
  \newcommand*\fsize{\dimexpr\f@size pt\relax}%
  \newcommand*\lineheight[1]{\fontsize{\fsize}{#1\fsize}\selectfont}%
  \ifx\svgwidth\undefined%
    \setlength{\unitlength}{2409.4488189bp}%
    \ifx\svgscale\undefined%
      \relax%
    \else%
      \setlength{\unitlength}{\unitlength * \real{\svgscale}}%
    \fi%
  \else%
    \setlength{\unitlength}{\svgwidth}%
  \fi%
  \global\let\svgwidth\undefined%
  \global\let\svgscale\undefined%
  \makeatother%
  \begin{picture}(1,0.64705882)%
    \lineheight{1}%
    \setlength\tabcolsep{0pt}%
    \put(0,0){\includegraphics[width=\unitlength,page=1]{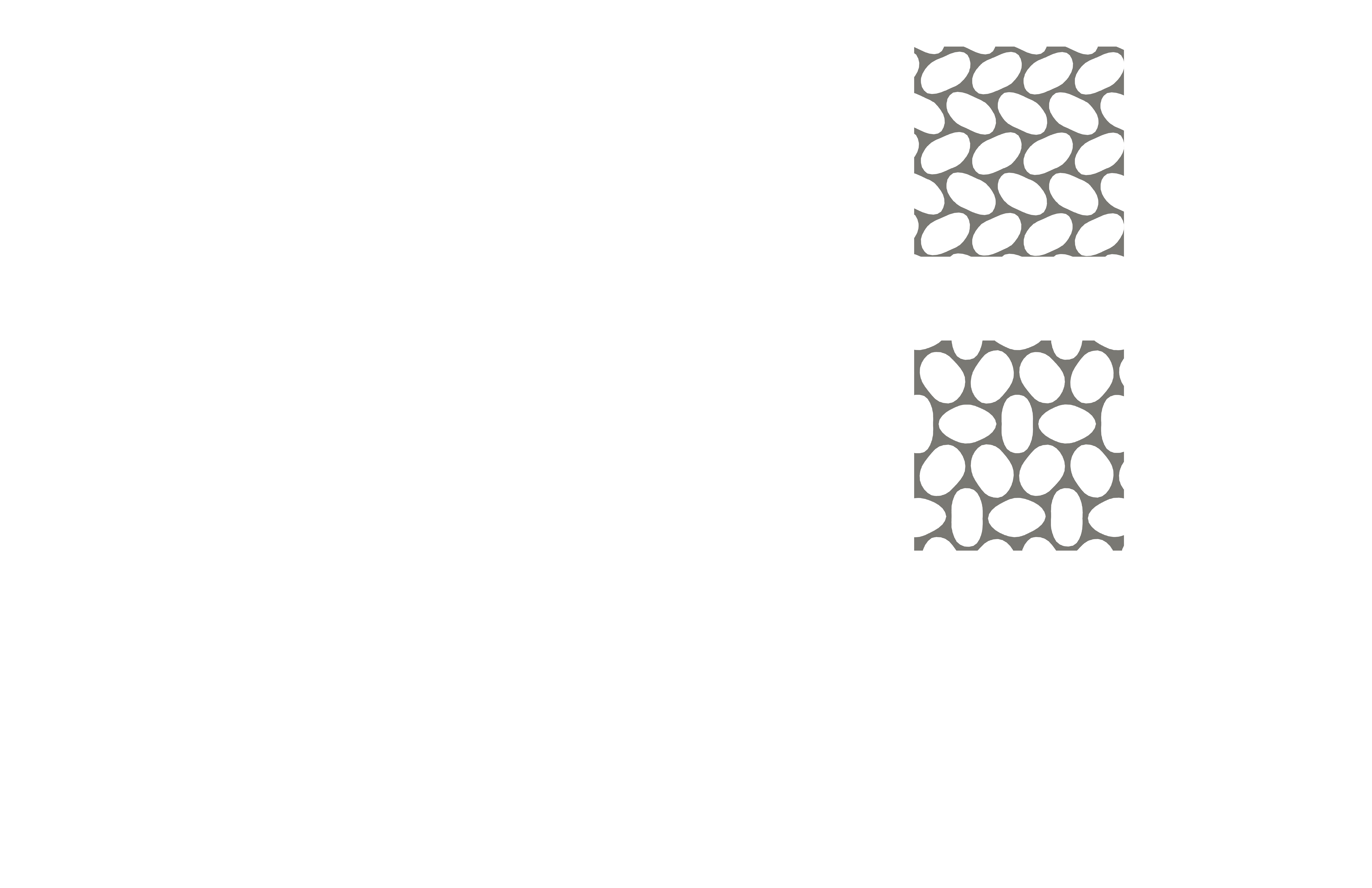}}%
    \put(0.85459343,0.52407569){\color[rgb]{0,0,0}\makebox(0,0)[lt]{\lineheight{1.25}\smash{\begin{tabular}[t]{l}\footnotesize pattern~I\end{tabular}}}}%
    \put(0.85459343,0.3139332){\color[rgb]{0,0,0}\makebox(0,0)[lt]{\lineheight{1.25}\smash{\begin{tabular}[t]{l}\footnotesize pattern~II\end{tabular}}}}%
    \put(0.85459343,0.09513286){\color[rgb]{0,0,0}\makebox(0,0)[lt]{\lineheight{1.25}\smash{\begin{tabular}[t]{l}\footnotesize pattern~III\end{tabular}}}}%
    \put(0,0){\includegraphics[width=\unitlength,page=2]{cruciform_deformed.pdf}}%
  \end{picture}%
\endgroup%
\label{cruciform-deformed}}
	\caption{Equi-biaxial compression with~$\varepsilon_{11}= \varepsilon_{22}=4\%$ of a cruciform geometry showing~(a) the reference geometry, and~(b) the deformed configuration including various patterns formed in different regions (cf. the zoomed-in images).}
	\label{results:cruciform-dns}
\end{figure}

As shown in Fig.~\ref{Figure.BL_1d}, complex boundary conditions may lead to spatial mixing of patterns. To test the ability of the proposed homogenization scheme to reproduce such a behaviour, in this example a macroscopic cruciform-like geometry with circular honeycomb microstructure is subjected to equi-biaxial compression with~$\varepsilon_{11}=\varepsilon_{22}=4\%$ applied via the four straight edges, cf. Fig.~\ref{cruciform}. The DNS solution, shown in Fig.~\ref{cruciform-deformed}, suggests that all the three patterns are triggered in this case (cf. zoomed-in images). Pattern~I can be observed in the top and bottom legs of the cruciform-like domain, while pattern~II is triggered mostly in the bulk of the geometry and extends towards either of the horizontal legs. The flower-like pattern~III is triggered slightly above and below the centre of the cruciform. The magnitudes of all three patterns are varying smoothly in space.

\begin{figure}[!t]
	\centering
	\begin{tabular}{ccc}
	\subfloat[discretization]{\includegraphics[scale=0.9]{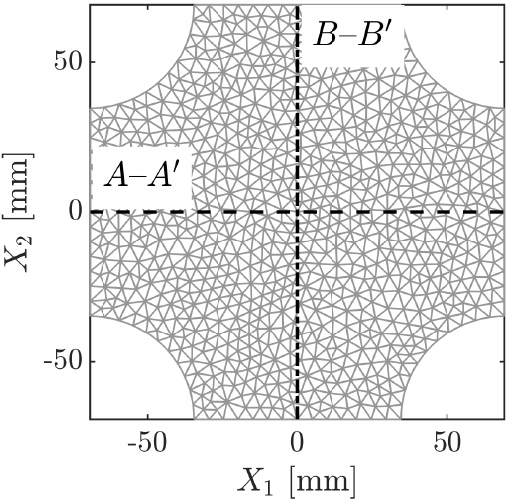}\label{results:cfa}} &
	\subfloat[$\vec{e}_1$ component of~$\vec{v}_0$ ~\mbox{[mm]}]{\includegraphics[scale=0.9]{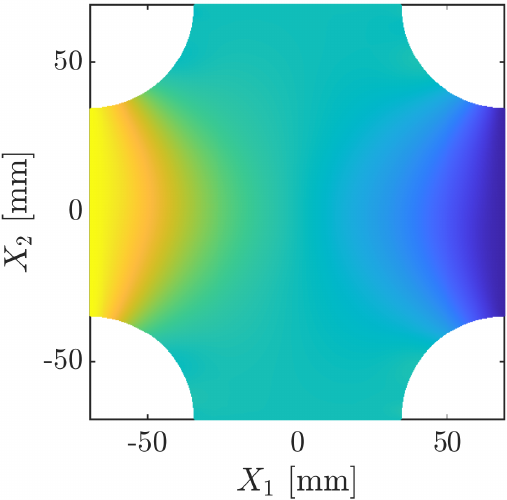}\label{results:cfb}} &
	\subfloat[$\vec{e}_2$ component of~$\vec{v}_0$~\mbox{[mm]}]{\includegraphics[scale=0.9]{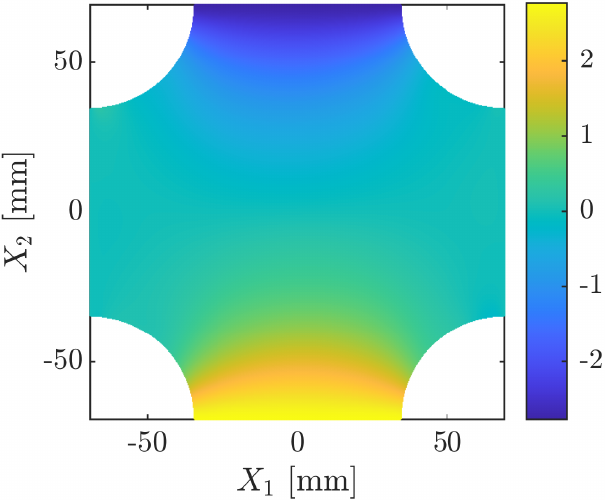}\label{results:cfc}}\\
	\subfloat[$v_1$~\mbox{[mm]}]{\includegraphics[scale=0.9]{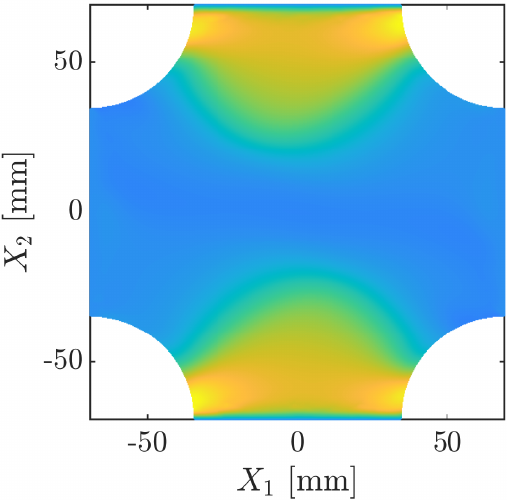}\label{results:cfd}} &
	\subfloat[$v_2$~\mbox{[mm]}]{\includegraphics[scale=0.9]{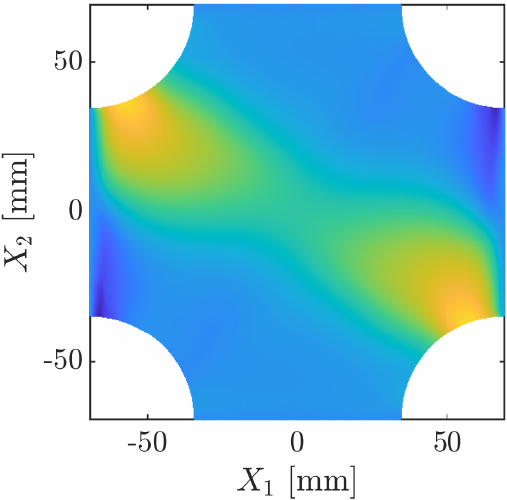}\label{results:cfe}} &
	\subfloat[$v_3$~\mbox{[mm]}]{\includegraphics[scale=0.9]{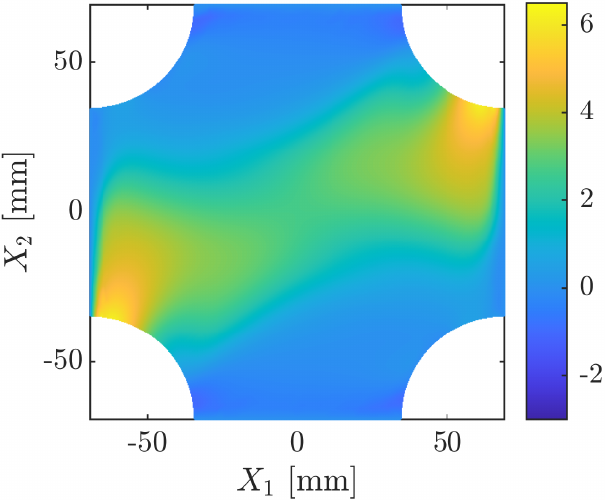}\label{results:cff}}
	\end{tabular}
	\caption{Results corresponding to the micromorphic homogenization obtained for the cruciform example of Fig.~\ref{cruciform} subjected to~$\varepsilon_{11}=\varepsilon_{22}=4\%$. (a)~Employed geometry and discretization. The two components of the macroscopic displacement field~$\vec{v}_0$ are shown in (b) horizontal and (c) vertical direction, whereas the distributions of the micromorphic fields~$v_i$ are shown in (d)--(f).}
	\label{results:cruciform-micromorphic}
\end{figure}

The micromorphic solution is obtained using quadratic isoparametric triangular elements with the three-point Gauss integration rule, with an excessively fine triangulation (shown in Fig.~\ref{results:cfa}) to make sure that macroscopically converged results are computed (note that the typical mesh element size is comparable to the adopted RVE size, i.e.~to~$2 \ell$). The solution was computed for one snapshot of the overall applied strain~$\varepsilon_{11}=\varepsilon_{22}=4\%$. The initial guess~$\vec{v}_0 = \vec{0}$~mm and~$v_i = 1$~mm was first adopted to compute a homogenized solution in which the microfluctuation field~$\vec{w}$ was neglected. This result was subsequently used as an initial guess for the full multiscale scheme accounting also for~$\vec{w}$. Although such a two-step initialization yields satisfactory results (as presented below), ideally a bifurcation analysis should be carried out in order to identify the proper macroscopic buckling modes. Such considerations might yield slightly different and more accurate results, but they lie outside of the scope of this contribution because a full macroscopic Newton solver with Hessian computation is required to this end, which will be reported separately.

\begin{figure}[!t]
	\centering
	\subfloat[section~$A$--$A'$, $X_2 = 0$~mm]{\includegraphics[scale=0.9]{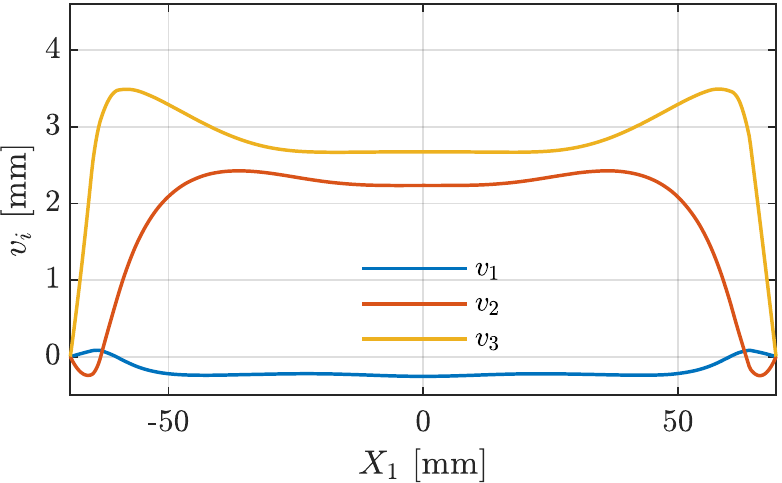}\label{cruciform:sectionsa}}
	\hspace{1em}
	\subfloat[section~$B$--$B'$, $X_1 = 0$~mm]{\includegraphics[scale=0.9]{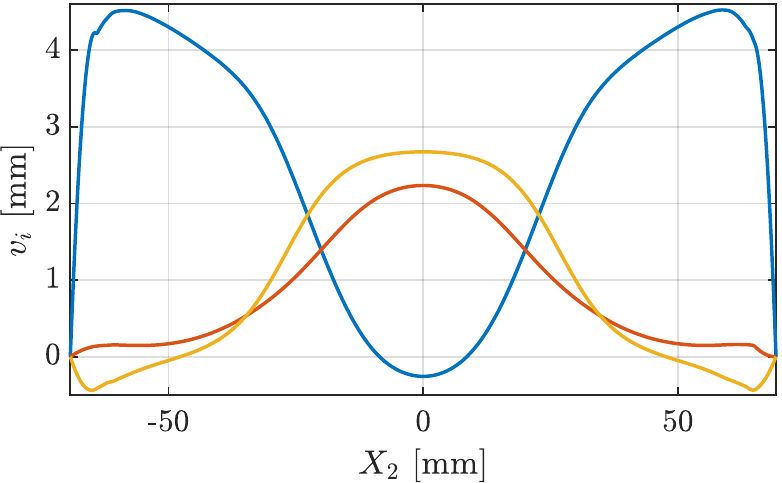}\label{cruciform:sectionsb}}
	\caption{Individual components of the micromorphic fields~$v_i$ along horizontal and vertical sections denoted as~$A$--$A'$ and~$B$--$B'$, as indicated in Fig.~\ref{results:cfa}. (a) Horizontal section~$A$--$A'$, along which mostly pattern~II is triggered. (b) Vertical section~$B$--$B'$, along which all three patterns are triggered.}
	\label{cruciform:sections}
\end{figure}
\begin{figure}[!t]
	\centering
	\subfloat[pattern~I]{
		\begin{tikzpicture}
		\node[rectangle,draw=red,line width=2pt,dash pattern=on 5pt off 2pt,inner sep=2pt] (I) at (0,0) {\includegraphics[height=3.5cm]{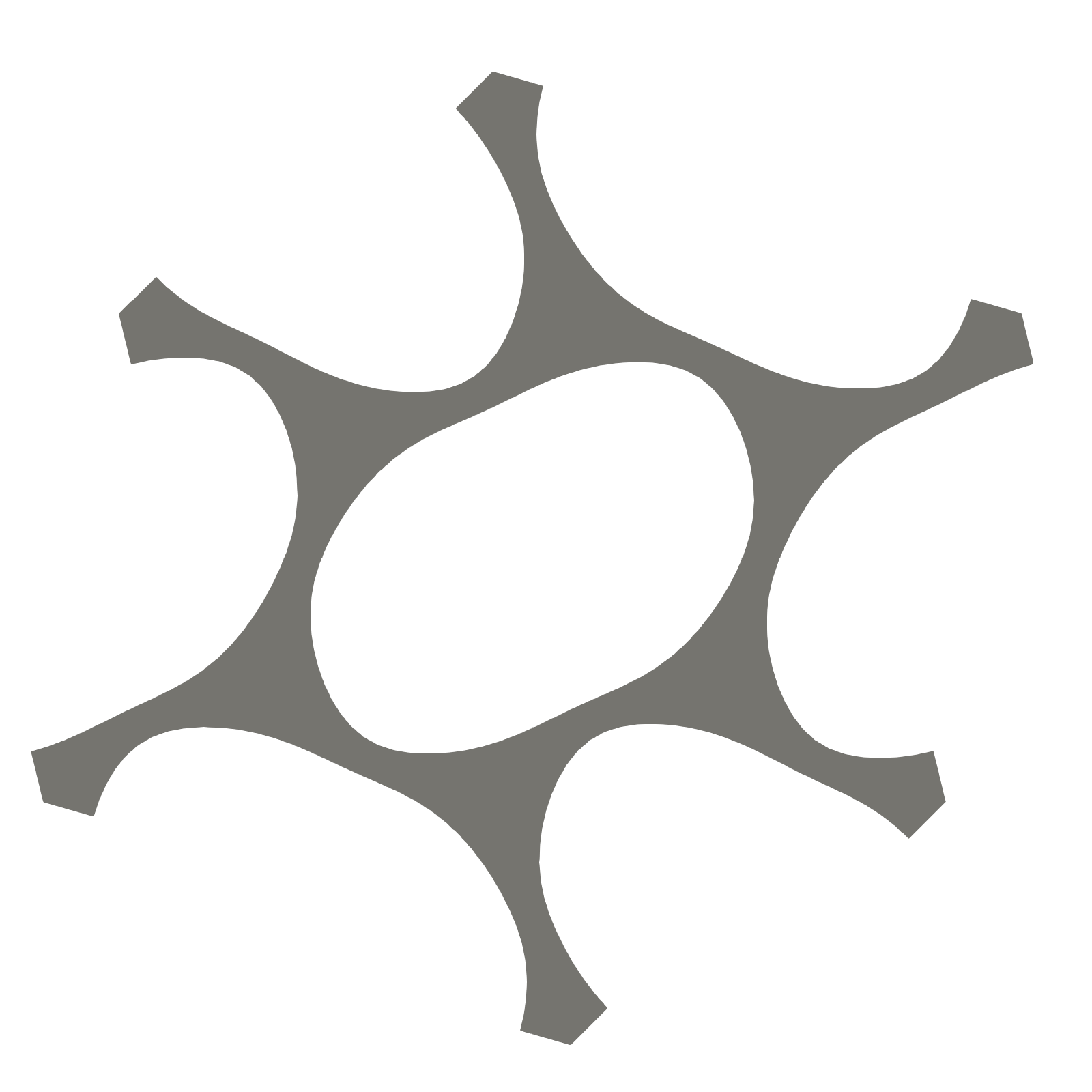}};
		\end{tikzpicture}
		\label{cruciform:rvesa}}
	\hspace{0.2em}
	\subfloat[pattern~II]{
		\begin{tikzpicture}
		\node[rectangle,draw=mygreen,line width=2pt,dash pattern=on 5pt off 2pt,inner sep=2pt] (II) at (0,0) {		\includegraphics[height=3.5cm]{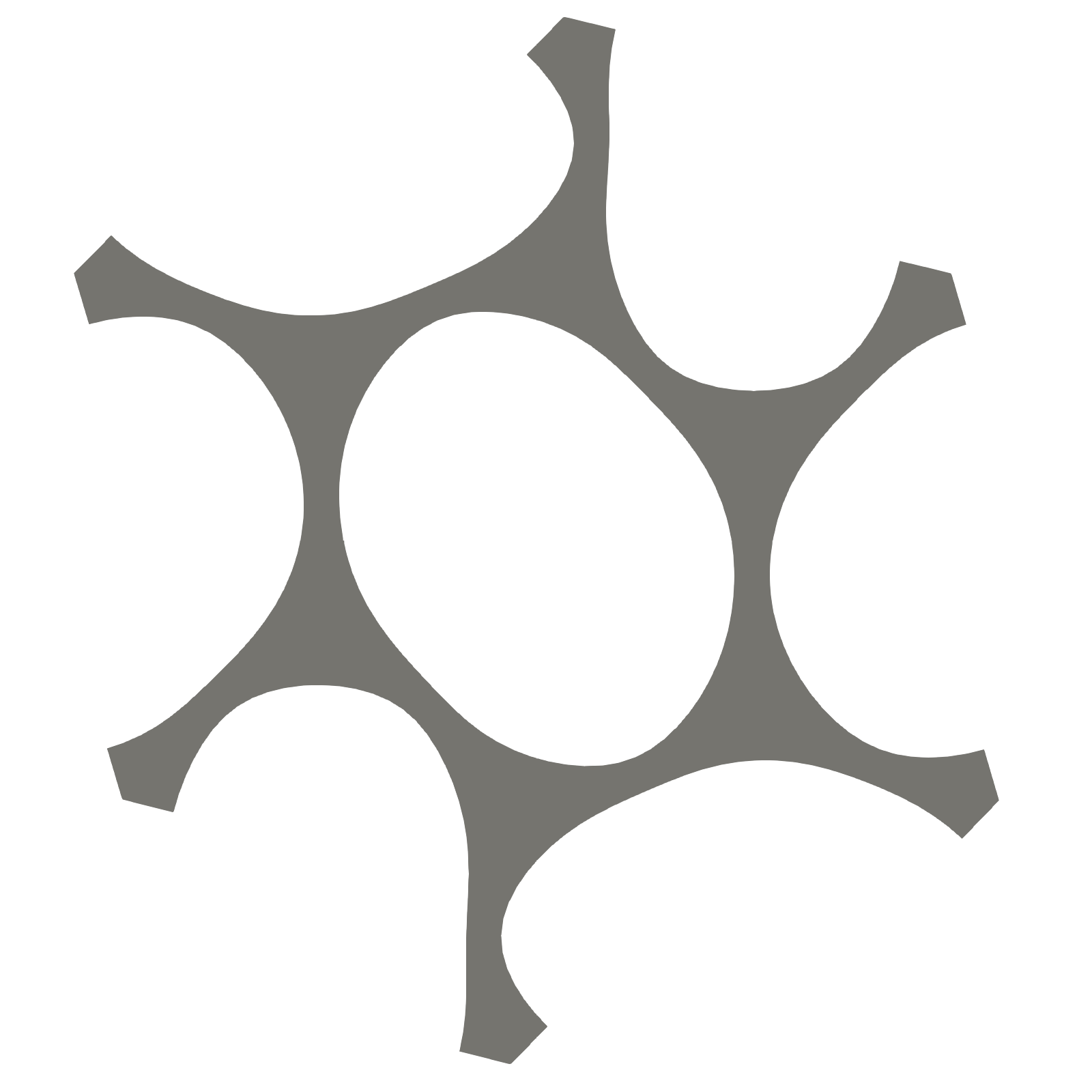}};
		\end{tikzpicture}		
		\label{cruciform:rvesb}}
	\hspace{0.2em}
	\subfloat[pattern~III]{
		\begin{tikzpicture}
		\node[rectangle,draw=blue,line width=2pt,dash pattern=on 5pt off 2pt,inner sep=2pt] (III) at (0,0) {\includegraphics[height=3.5cm]{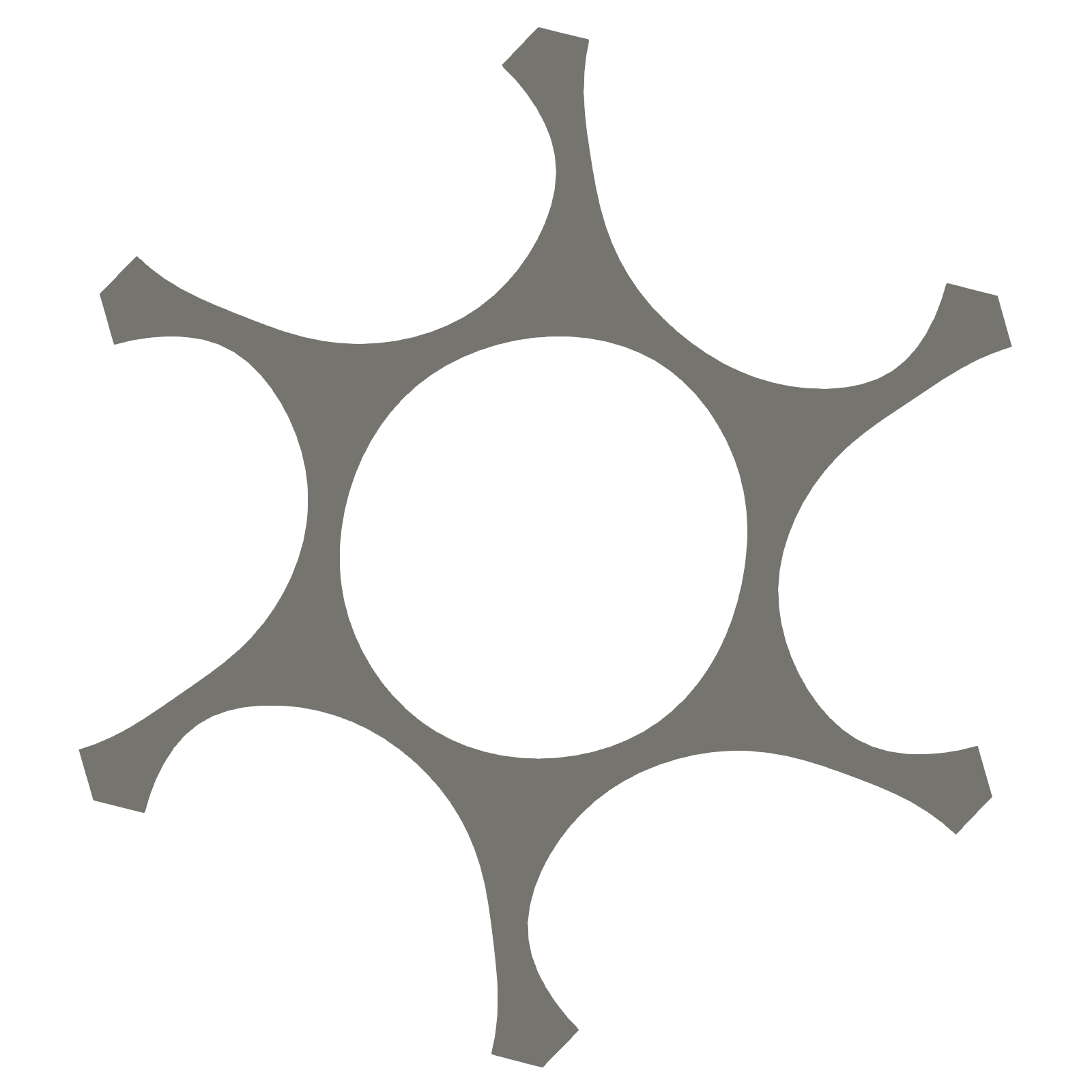}};
		\end{tikzpicture}			
		\label{cruciform:rvesc}}
	\caption{Deformed RVEs of the micromorphic scheme corresponding to spatial positioning of individual patterns depicted in Fig.~\ref{cruciform-deformed} (the zoomed-in images). (a) Pattern~I, (b) pattern~II, and~(c) pattern~III.}
	\label{cruciform:rves}
\end{figure}

The obtained micromorphic results are summarized in Fig.~\ref{results:cruciform-micromorphic}, where we first notice that the individual components of the~$\vec{v}_0$ field possess the proper symmetries, respecting the considered geometry and applied loading. Second, the micromorphic fields vary smoothly throughout the specimen domain, thus resulting in various patterns. From Fig.~\ref{results:cfd} we further observe that mode~I (i.e. pattern~I) is triggered mostly close to the top and bottom horizontal boundaries of the specimen, whereas modes~II and~III localize in the two diagonals across the specimen domain. Such observations are in agreement with the DNS results of Fig.~\ref{cruciform-deformed}.

For better clarity, Fig.~\ref{cruciform:sections} shows the resulting micromorphic fields~$v_i$ along two sections (horizontal~$A$--$A'$ and vertical~$B$--$B'$), indicated in Fig.~\ref{results:cfa} by the dashed and dash-dotted lines. From Fig.~\ref{cruciform:sectionsa} it may be concluded that along the entire horizontal section~$A$--$A'$, pattern~II is triggered ($v_1 \approx 0$~mm, whereas~$v_2 \approx v_3 \neq 0$~mm). Along the vertical section~$B$--$B'$ (Fig.~\ref{cruciform:sectionsb}), however, mostly pattern~I is triggered close to the top and bottom boundaries, which transforms through pattern~III ($v_1 \approx v_2 \approx v_3$) to pattern~II, localized in agreement with Fig.~\ref{cruciform:sectionsa} at~$X_2 \approx 0$~mm. The observed behaviour also correlates satisfactorily with the DNS results shown in Fig.~\ref{cruciform-deformed}, although pattern~II seems to significantly vary its magnitude along the~$A$--$A'$ section in the DNS results whereas it is almost constant in the micromorphic scheme. Fig.~\ref{cruciform:rves} finally shows three deformed RVEs, with spatial positioning matched to the three coloured squares in the DNS results (i.e.~red, green, and blue) shown in Fig.~\ref{cruciform-deformed}. The first RVE (Fig.~\ref{cruciform:rvesa}) corresponds to magnitudes of individual micromorphic fields~$v_1 = 4.5$~mm, $v_2 = 0.2$~mm, $v_3 = -0.3$~mm, i.e.~distorted pattern~I, and its deformed shape compares well with the DNS result of Fig.~\ref{cruciform-deformed}. The second RVE~(Fig.~\ref{cruciform:rvesb}), corresponds to~$v_1 = -0.2$~mm, $v_2 = 2.5$~mm, $v_3 = 2.9$~mm, i.e.~to distorted pattern~II, again matching well with the patterning of the DNS results. For the last RVE (Fig.~\ref{cruciform:rvesc}), magnitudes of the three micromorphic fields are~$v_1 = 1.9$~mm, $v_2 = 1.2$~mm, $v_3 = 1.8$~mm, suggesting that a distorted pattern~III is triggered. Since the DNS result shows a smaller patterning magnitude, agreement between the two deformed shapes is less obvious in this case, although pattern~III can be resembled in both as well.

The computed reaction forces exerted by the cruciform specimen are overall approximately~$6\%$ softer for the micromorphic homogenization scheme compared to the DNS. In particular, the vertical reaction force is $-449.1$~kN and $-478.9$~kN for the micromorphic and DNS result, respectively, whereas the horizontal reaction force is $-457.7$~kN and $-475.6$~kN for the micromorphic and DNS result.
%
%
\section{Summary and Conclusions}
\label{summary}
In this paper, an extended micromorphic homogenization framework has been presented, capable of capturing multiple pattern transformations emerging in elastomeric mechanical metamaterials. Based on the approximate global energy governing the evolution of the multiscale system, the associated Euler--Lagrange equations have been derived and discussed. The resulting system of partial differential equations govern the equilibrium of a micromorphic continuum with multiple micromorphic fields representing the magnitudes of long-range correlated patterning fields.

The main results of this paper can be summarized as follows:
\begin{enumerate}
	
	\item An overview of pattern transformations emerging in elastomeric honeycombs has been provided, and an appropriate choice of patterning modes has been incorporated into the proposed kinematic ansatz.

	\item Upon making use of the variational homogenization framework, effective governing equations enriched with the magnitudes of the underlying patterning fluctuation fields have been derived. The additional micromorphic fields reveal the magnitudes of the individual modes, reflecting the kinematic interaction between adjacent cells, essential for capturing size effects. Although size effects were not in the focus of this contribution, and hence not discussed in detail, the results obtained for the cruciform-shaped specimen are indeed size dependent. 
	
	\item It has been demonstrated that the proposed methodology is capable of capturing the temporal switching of patterns exhibited by the studied honeycomb-like microstructure, where a periodic unit cell shows two bifurcation points at two applied strain levels for a fixed biaxiality ratio.

	\item To investigate spatial mixing of patterns, a specimen with cruciform-like geometry subjected to equi-biaxial compression was studied. It was shown that the proposed homogenization framework is capable of capturing the spatial mixing adequately. In terms of reaction forces, the micromorphic scheme is approximately~$6\%$ softer compared to the corresponding DNS results.
	
	\item Although satisfactory results were obtained with ad hoc initial guesses, the derivation of a full Newton solver including macroscopic Hessians is required in order to allow for proper bifurcation analyses and an efficient solution scheme.
	
\end{enumerate}

The presented micromorphic computational scheme thus holds promise for the efficient prediction of the overall mechanical behaviour of elastomeric microstructures with multiplicity of patterning deformation.
%
%
%
%
\section*{Acknowledgements}
The research leading to these results has received funding from the European Research Council under the European Union's Seventh Framework Programme (FP7/2007-2013)/ERC grant agreement \textnumero~[339392].
%
%
\bibliography{mybibfile}
\end{document}